\documentclass[twoside,a4paper,12pt,reqno]{amsart}
\usepackage{latexsym}
\usepackage{amsmath}
\usepackage{amssymb}
\usepackage{amsthm}
\usepackage{amsxtra}
\usepackage{amsfonts}
\usepackage{mathrsfs}
\usepackage{amsbsy}

\newtheorem{teo}{Theorem}[section]

\newtheorem{lem}[teo]{Lemma}

\newtheorem{prop}[teo]{Proposition}

\newenvironment{dem}{\vspace{.2cm}\noindent {\bf Proof }\\}{\newline \hspace{1cm} \hspace{1cm}\flushright \hfill $\square$ \newline}

\newcommand{\II}{\mathcal I}

\newcommand{\n}{\noindent}

\newcommand{\ve}{\varepsilon}
\newcommand{\erre}{\mathbb{R}} 

\newcommand{\enne}{\mathbb{N} }

\newcommand{\de}{\delta} 
\newcommand{\al}{\alpha}

\newcommand{\f}{\frac}
\newcommand{\ba}{\begin{eqnarray}} \newcommand{\ea}{\end{eqnarray}}
\newcommand{\be}{\begin{equation}} \newcommand{\ee}{\end{equation}}
\newcommand{\bdm}{\begin{displaymath}} \newcommand{\edm}{\end{displaymath}} 
\newcommand{\brr}{\begin{array}}\newcommand{\err}{\end{array}}

\newcommand{\lf}{\left}
\newcommand{\ri}{\right}

\newcommand{\bml}{\begin{gather}} 
\newcommand{\eml}{\end{gather}}

\newcommand{\vs}{\vspace{.5cm}}
\newcommand{\bo}{\boldsymbol}
\newcommand{\ran}{\rangle}
\newcommand{\lan}{\langle}
\newcommand{\bs}{\boldsymbol}
\newcommand{\bsx}{\boldsymbol{\xi}}
\newcommand{\bsz}{\boldsymbol{z}}
\newcommand{\mis}{e^{\f{i}{\ve}n\tau}}

\numberwithin{equation}{section}

\begin{document}


\vs

\title[ ]{ASYMPTOTIC EXPANSION FOR THE WAVE FUNCTION IN A ONE-DIMENSIONAL MODEL OF INELASTIC INTERACTION}

\author{Domenico Finco}

\address{Finco: Dipartimento di Matematica, Sapienza  Universit\'a di Roma }

\curraddr{P.le Aldo Moro, 2 - 00185 Roma,  Italy}

\email{finco@mat.uniroma1.it}

\author{Alessandro Teta} 
\address{ Teta: Dipartimento di Matematica Pura ed Applicata, Universit\`a di L'Aquila}

\curraddr{Via Vetoio - Loc. Coppito - 67010 L'Aquila, Italy}

\email{teta@univaq.it}

{\maketitle}

\vs

\begin{abstract}
We consider a two-body quantum system in dimension one composed by a test particle interacting with an harmonic oscillator placed at the position $a>0$. At time zero the test particle is concentrated around the position $R_0$ with average velocity $\pm v_0$ while the oscillator is in its ground state. In a suitable scaling limit, corresponding for the test particle to a semi-classical regime with small energy exchange with the oscillator, we give a complete asymptotic expansion of the wave function of the system in both cases $R_0 <a$ and $R_0 >a$.

\end{abstract}

\vs
\vs

\section{Introduction}

\vs

\n
The analysis of the classical behavior emerging in a quantum system is a widely studied subject both from the theoretical and from the experimental point of view. It is generally accepted that such behavior cannot be explained by simply taking the limit $\hbar \rightarrow 0$ for the isolated system but the crucial role of the environment must be taken into account. In particular the environment is responsible for  the suppression of the quantum interference between two different components of a superposition state of the system. The dynamical mechanism producing the suppression is known as decoherence (\cite{bgjks}, \cite{gjkksz}, for some rigorous results see also \cite{afft}, \cite{ccf}, \cite{d}, \cite{duft}, \cite{ds}).

\n
In this context, a particularly important problem is the explanation of the appearance of straight tracks in a cloud chamber produced by an $\alpha$-particle emitted in the form of a spherical wave. The problem was raised by Mott (\cite{m}) who gave an heuristic explanation based on the analysis of the interaction of the $\alpha$-particle with only two atoms of the vapour. In \cite{dft1} and \cite{dft2} the same problem was studied in dimension one and three, and a rigorous result was given up to second order in perturbation theory. In this paper we  consider an even simpler situation, i.e. a test particle in dimension one interacting with one harmonic oscillator, and we want to give a complete asymptotic expansion of the wave function of the system valid  at any order in perturbation theory.

\n
We want to emphasize that a complete and detailed analysis of this case is the starting point for the analysis
of the more realistic model of an $\alpha$-particle interacting with many atoms.

\n
The Schr\"{o}dinger equation for the wave function $\Psi(t)$  of the system  is 

\be\label{se}
i \hbar \f{\partial \Psi(t)}{\partial t} = -\f{\hbar^2}{2M} \Delta_R \Psi(t)  -\f{\hbar^2}{2m} \Delta_r \Psi(t) + \f{1}{2} m \omega^2 (r-a)^2 \Psi(t) + \lambda V(\delta^{-1}(R-r)) \Psi(t)
\ee

\n
In (\ref{se}) we have denoted by $R$, $M$ the position coordinate and  the mass of the test particle, and by $r$,   $m$,  $\omega$ the position coordinate, the mass and the frequency of the oscillator placed at the position $a>0$. The smooth interaction potential is denoted by $V$ and $\lambda >0$, $\delta >0$ are the coupling constant and the effective range of the interaction.  

\n
We consider two possible initial state  chosen in the product form

\be
\Psi^{\pm}_0 (R,r)= \f{1}{\sqrt{\sigma}} \, e^{\pm i{\f{M v_0}{\hbar}R}} \,\eta (\sigma^{-1}(R-R_0))\, \varphi_0 (r)
\ee
where $\sigma, v_0  >0$,  $R_0 \in \erre$, $\eta \in \mathcal{S}(\erre)$, $\|\eta\|_{L^2}=1$, where $\mathcal{S}(\erre)$
is the Schwartz space. We have denoted by   $\pm v_0,  R_0 $ the average  velocity and position of the test particle at time zero and  by $\varphi_0$  the ground state of the oscillator. More generally,   eigenfunctions and eigenvalues of the oscillator are denoted  by

\be
\varphi_n (r) = \f{1}{\sqrt{\gamma}} \phi_n (\gamma^{-1}(r-a)) , \;\;\;\;\; n \in \enne , \;\;\;\;\; \gamma = \sqrt{\f{\hbar}{m \omega}}
\ee
\be
E_n = \hbar \omega \left(n + \f{1}{2} \right)
\ee

\n
where $\phi_n$ is the Hermite function of order $n$.

\n
We want to study the asymptotic behavior of the wave function of the system under a suitably chosen scaling limit. More precisely we introduce a small parameter $\varepsilon >0$ and consider a semi-classical scaling for the test particle for $\varepsilon \rightarrow 0$, i.e.
\be
\hbar = \ve^2 \qquad M=1 \qquad 
\sigma= \ve
\ee

\n
For the physical parameters of the oscillator we fix 
\be
m=\ve \qquad \omega = \ve^{-1}
\ee

\n
which imply  $\gamma =\ve$, i.e.  the oscillator is well localized around the position $a$, and  $E_n=\ve (n + 1/2)$, i.e. the energy level spacing of the oscillator is much smaller than the kinetic energy of the test particle.  

\n
Finally we choose the range of the interaction of the same order of the initial localization of test particle and oscillator  and fix the coupling constant such that the deviation from the free dynamics is small, i.e.
\be
\delta =\ve \qquad \lambda =\ve^2
\ee

\n
Under the above scaling, the system is  described by the rescaled wave function $\Psi_{\ve}^{\pm} (t)$ given by

\ba\label{psir}
&&\Psi_{\ve}^{\pm} (t) = e^{-i\f{t}{\ve^2} H^{\ve}} \Psi_{\ve,0}^{\pm}
\ea
where $H^{\ve}$ is the Hamiltonian of the system
\ba\label{hri}
&&H^{\ve}=H_0^{\ve} +\ve^2 \, V^{\ve}\\
&&H_0^{\ve}= h_0^{\ve}+ h_a^{\ve}\label{h0r} \equiv -\f{\ve^4}{2} \Delta_R + \f{1}{\ve} \left[ - \f{\ve^4}{2} \Delta_r + \f{1}{2} (r-a)^2 \right] \\
&&V^{\ve}= V(\ve^{-1}(R-r))\label{vr}
\ea
and $\Psi^{\pm}_{\ve,0}$ are the two possible initial states
\ba\label{isr}
&&\Psi^{\pm}_{\ve,0}(R,r)=\psi^{\pm}_{\ve} (R) \, \phi_0^{\ve} (r) \\
&&\psi^{\pm}_{\ve} (R)=  \f{1}{\sqrt{\ve}} \, e^{\pm i \f{v_0}{\ve^2} R }\,   \eta(\ve^{-1}(R-R_0)) \,\\
&&\phi_0^{\ve} (r)=\f{1}{\sqrt{\ve}}\phi_0 (\ve^{-1}(r-a))
\ea

\n
The aim of this paper is to characterize the asymptotic expansion of the wave function of the system $\Psi^{\pm}_{\ve}(t)$ for $\ve \rightarrow 0$.

\n
We list some notation which will be used in the following.
\be
\bo{a} = (a_1,\ldots,a_l) , \;\;\;\;\;\; d\bo{a}=da_1\ldots da_l
\ee
\be
|\bs a| = \sum_{j=1}^l |a_j| \qquad \qquad  \| \bs a \|^2 = \sum_{j=1}^l |a_j|^2 \qquad \qquad
\bs a \cdot \bs b = \sum_{j=1}^l a_j b_j
\ee
The norm of $L^p(\erre)$ will be denoted by $\| \cdot \|_{L^p}$ and the inner product of $L^2(\erre)$ will 
be denoted by by $\lan \cdot , \cdot \ran$.
The norm of the total Hilbert space $L^2(\erre^2)$ will be simply denoted by $\| \cdot \|$.
Moreover we introduce the norm
\be
||| f |||_{k,p} = \sum_{j=0}^k \| f^j \,\lan \cdot \ran^{k-j} \|_{L^p }
\ee
where $f^j$ denotes the derivative of $f$ of order $j$ and $\lan x \ran =(1+x^2)^{1/2}$. The symbol $\hat{f}$ indicates the Fourier transform of $f$. 
The propagator of the  harmonic oscillator centered in the origin and corresponding to $\hbar=m=\omega=1$ will be denoted by $U(t)$. Its integral kernel is explicitly given by
\be
U(t; x,y) =
\sum_{n=0}^{\infty} e^{-i(n+\f 1 2 )t} \phi_n (x) \phi_n (y) =
\f{1}{[\pi( 1-e^{-2it})]^{1/2} } 
e^{-\f{x^2 + y^2}{2 i \tan t} +\f{xy}{i \sin t} }
\ee
Finally $c$ will be a positive numerical constant whose value may change line by line.

\n
The paper is organized as follows: in section 2 we state our results and outline the strategy of the proof.
The proofs are postponed to section 3 while in section 4
we explicitly compute the first terms of the asymptotic expansion and discuss a physical application.
In the appendix we give the proof of proposition \ref{prop1}.

\vs

\section{Results}
\n
We first give some heuristic arguments in order to explain the qualitative behavior of the system. 
Let us consider the initial state $\Psi^{+}_{\ve,0}$, i.e. the oscillator in its ground state and the test particle  initially  described by a coherent state with average position $R_0$ and positive velocity $v_0$.  For small  $t>0$ the effect of the interaction is negligible and then the  state of the test particle will undergo a free semi-classical evolution.  As the time increases,  it is reasonable to expect two completely different situations  in the cases $R_0 <a$ (the test particle  "hits"  the oscillator) and $R_0>a$ (the test particle does not "hit" the oscillator).  

\n
In the case $R_0<a$, taking into account that the effective dimension of the oscillator and the range of the interaction are of order $\ve$, one has that the interaction will take place approximately at the (classical) impact time  $\tau$ given by

\be\label{imt}
\tau = \f{|R_0 -a|}{v_0}
\ee

\n
Due to the interaction, the state of the test particle and of the oscillator will be almost instantaneously modified. Then for $t>\tau$ we shall find the oscillator in a superposition of  stationary states while  for the test particle we expect to find  again a free semi-classical propagation, but of a modified coherent state.  Therefore for this case the problem is reduced to  compute  the modification of the state of the system determined by the interaction at $t=\tau$. This will be done using a perturbative expansion of the wave function of the system and exploiting stationary phase methods to control each term of the expansion. Therefore we shall refer  to this case as the stationary case.

\n
In the case $R_0>a$ the situation is easier since we have only to verify that for any $t>0$ the state of the system has an essentially unperturbed evolution, with the oscillator in its ground state and the test particle evolving according to a free semi-classical propagation of its initial state. This will be done using standard non-stationary phase methods and therefore this case will be referred to as the  non-stationary case.

\n
It is obviously true that if we consider  the other possible initial state $\Psi^{-}_{\ve,0}$ then the opposite situation occurs, i.e. the case $R_0<a$ corresponds to the non-stationary case while the case $R_0>a$ corresponds to the stationary case.

\n
Let us outline the strategy of the proof for the characterization of the asymptotic behavior of
 $\Psi^+_{\ve}(t)$ (for $\Psi^-_{\ve}(t)$ the analysis is completely analogous).

\n
The starting point is to represent the solution using Duhamel's formula
\ba\label{du}
&&\Psi^{+}_{\ve}(t)=e^{-i\f{t}{\ve^2} H_0^{\ve}} \Psi^+_{\ve,0}  -i  \int_0^t \!\! ds\, e^{-i  \f{t-s}{\ve^2} H^{\ve}} V^{\ve} e^{-i \f{s}{\ve^2} H^{\ve}_0} \Psi^+_{\ve,0}
\ea

\n
After $k+1$ iterations we get

\be\label{dui}
\Psi^{+}_{\ve}(t)=e^{-i\f{t}{\ve^2} H_0^{\ve}} \left( \Psi^+_{\ve,0}  +  \sum_{l=1}^{k+1}     I^{\ve}_l (t) 
\right)  +\mathcal{R}^{\ve}_{k+1}(t)
\ee

\n
where we have denoted for any $l=1,2\ldots , k+1$

\be\label{il}
I^{\ve}_l (t)\equiv (-i)^l \int_0^t \!\! ds_l \ldots \int_0^{s_{2}} \!\!\! \!\!ds_1 \, 
 e^{i  \f{s_l}{\ve^2} H_0^{\ve}} \, V^{\ve} \, e^{-i \f{s_l}{\ve^2} H^{\ve}_0}
\ldots
e^{i  \f{s_1}{\ve^2} H_0^{\ve}} \, V^{\ve} \, e^{-i \f{s_1}{\ve^2} H^{\ve}_0} \Psi_{\ve,0}^+
\ee
and the rest $\mathcal{R}^{\ve}_{k+1}(t)$ is 
\be
\label{re}
\mathcal{R}^{\ve}_{k+1}(t) = - i \!\int_0^t \!\! ds
\,   e^{-i  \f{t-s}{\ve^2} H^{\ve}} V^{\ve} e^{-i \f{s}{\ve^2} H^{\ve}_0}    \,
I^{\ve}_{k+1} (s)   
\ee

\n
It is convenient to consider the expansion of the wave function $\Psi^{+}_{\ve}(t)$ on the basis of eigenfunctions of the oscillator, i.e.

\ba\label{ex}
&&\Psi^+_{\ve}(t; R,r) = \sum_{n=0}^{\infty} f_n^{\ve} (t;R) \, \phi_n^{\ve} (r)\\
&&f_n^{\ve} (t;R)= \, \lan \phi_n^{\ve}\, , \Psi^+_{\ve}(t;R,\cdot) \ran\label{cf}
\ea
and 
\ba\label{par}
&&\|\Psi^+_{\ve} (t) \|^2 = \sum_{n=0}^{\infty} \|f_n^{\ve} (t) \|^2_{L^2}
\ea

\n
From (\ref{dui}) and (\ref{cf})  we have

\be
f_n^{\ve} (t) =e^{-i\f{t}{\ve^2} h_0^{\ve}} \left( \! \delta_{n0} \, e^{-i \f{t}{2\ve}}  \psi^+_{\ve} + e^{-i \f{t}{\ve} (n+\f{1}{2})}     \sum_{l=1}^{k+1} \lan  \phi_n^{\ve} \, , I_l^{\ve} (t) \ran \! \right)  + \lan \phi_n^{\ve} \, , \mathcal{R}^{\ve}_k (t) \ran
\ee

\n
The next step is to give  an explicit expression for the quantity $\lan \phi_n^{\ve} \, , I_l^{\ve}(t) \ran$.  
We introduce the rescaled variable $x$ given by
\be 
x=\ve^{-1}(R-R_0)
\ee
and the following phase function
\be
\varphi_{\ve} (x) = \f{v_0}{\ve} x - \f{v_0}{\ve^2}R_0
\ee
We have the following representation formula.

\begin{prop}\label{prop1}
For any $\ve >0$ and $n,l \in \enne$ we have 
\be\label{pnil}
\lan \phi_n^{\ve} \,, I_l^{\ve}(t; \ve x + R_0, \cdot) \ran \!
= \! \f{1}{\sqrt{\ve}} e^{i\varphi_{\ve} (x)} (-i)^l \!\!\! \sum_{n_1 \ldots n_l=0}^{\infty}\! \!\!\delta_{n_1 0} \!
\int_0^t \!\! \!ds_l \!\! \int_0^{s_l} \!\!\! ds_{l-1} \ldots \!\! \int_0^{s_2} \!\!\!\! ds_1 \!\! \int \!\! d\bo{\xi} \, F_{\bo{n}}(\bo{s}, \bo{\xi} ;x) e^{\f{i}{\ve} \Phi_{\bo{n}} (\bo{s},\bo{\xi})}
\ee
where 
\be\label{fnn}
F_{\bo{n}}(\bo{s},\bo{\xi};x)= \eta(x-\bo{s}\! \cdot \! \bo{\xi}) \prod_{j=1}^{l} 
\hat{V}(\xi_j) \widehat{\phi_{n_{j+1}} \phi_{n_j}} (\xi_j) 
e^{i \left( \f{s_j}{2} \xi_j^2 +  \xi_{j-1} \sum_{k=j}^{l} \xi_k  s_k - \xi_j  x \right)}
\ee

\ba\label{fin}
&&\Phi_{\bo{n}}(\bo{s}, \bo{\xi}) = \sum_{j=1}^{l} \Phi_j (s_j,\xi_j) = \sum_{j=1}^{l} \big[ (a-R_0) \xi_j + (n_{j+1}-n_j) s_j - v_0 s_j \xi_j \big] 
\ea
and $n_{l+1} \equiv n$, $\xi_0 \equiv 0$.
\end{prop}

\vs
\n
The proof is a long but straightforward computation and it is postponed to the appendix. From formula (\ref{pnil}) one sees that the problem is reduced to the analysis of a highly oscillatory integral. The asymptotic behavior for $\ve \rightarrow 0$ is then characterized by the presence (or the absence) of critical points of the phase $\Phi_{\bo{n}}$ in the integration region.  The unique critical point of the phase is easily found

\ba\label{sc}
&&\bo{s^c} = sgn \, (a-R_0) \, \bo{\tau} , \;\;\;\;\;\; \bo{\tau}\equiv (\tau, \ldots , \tau) 
\\
&&\bo{\xi^c}=(\xi_1^c,\ldots ,\xi_l^c), \;\;\;\;\;\; \xi_j^c = \f{n_{j+1}-n_j}{v_0} , \;\;\;\;\;\; j=1,\ldots , l \label{xic}
\ea

\n
For $R_0 >a$ the critical point lies outside the integration region and then we can use standard non-stationary phase arguments to show that the contribution of the integral can be made arbitrarily small. The result is summarized in the following theorem.

\begin{teo}\label{t2}
Let us fix  $k \in \enne$, $t>0$ and assume that $||| V|||_{k+1, 1} < \infty$.

\n
If $R_0>a$ then 
\ba\label{>+}
&&\Psi^{+}_{\ve} (t) = e^{-i\f{t}{\ve^2} H_0^{\ve}}  \Psi^{+}_{\ve,0}  + \mathcal{N}^{+}_{\ve}(t)
\ea
and if $R_0<a$ then 
\ba\label{<-}
&&\Psi^{-}_{\ve} (t) = e^{-i\f{t}{\ve^2} H_0^{\ve}}  \Psi^{-}_{\ve,0}  + \mathcal{N}^{-}_{\ve}(t)
\ea
The remainders  $\mathcal{N}^{\pm}_{\ve }(t)$ 
satisfy
\ba
&&\| \mathcal{N}^{\pm}_{\ve}(t) \| < D_k\, \ve^{k+1}
\ea
where $D_k$ depends on  $t$, $v_0$, $V$, $\eta$.

\end{teo}

\n
For $R_0<a$ the critical point lies  on a one dimensional subset of the boundary of the integration region. This means that the explicit computation  of the asymptotic expansion in power of $\ve$ is rather involved (see  e.g. \cite{bh},\cite{f},\cite{hor}).  
In order to simplify the analysis, we shall  exploit the fact that the phase in formula (\ref{pnil}) is linear in the variable $\bo{\xi}$. 
More precisely,  we have

\be\label{fuf1}
\int \!\! d\bo{\xi} \, F_{\bo{n}}(\bo{s}, \bo{\xi};x) e^{-\f{i}{\ve} \Phi_{\bo{n}} (\bo{s}, \bo{\xi}) } 
=
e^{ \f{i}{\ve} v_0 \bo{\xi^c} \cdot \bo{s}}
\int d\bsx F_{\bs n}(\bs s , \bsx ; x) e^{\f{i}{\ve} v_0(\bo{s} -\bo{\tau}) \cdot \bsx }
\ee
\n
where we have used the notation introduced in (\ref{sc}), (\ref{xic}).  Exploiting   (\ref{fuf1}) in (\ref{pnil}),  introducing the new integration variables $z_j=\ve^{-1} v_0 (s_j  - \tau)$, $j=1, \ldots ,l$, and defining
\be\label{omta}
\Omega_{\ve} \equiv \{\bo{z} \in \erre^l \;|\; -\ve^{-1}v_0 \tau < z_1 <z_2 <\ldots <z_l <\ve^{-1}v_0(t-\tau) \} 
\ee
we get

\be\label{pnil2}
\lan \phi_n^{\ve} \,, I_l^{\ve}(t; \ve x + R_0, \cdot) \ran \!
=\! \left( \! \! \f{ \ve}{i v_0} \!\right)^{\!\!l}  
\f{1}{\sqrt{\ve}} e^{i\varphi_{\ve} (x)}\,  e^{\f{i}{\ve} n \tau} \!\! \!\! \sum_{n_1 \ldots n_l=0}^{\infty}\! \!\!\delta_{n_1 0} \! \int_{\Omega_{\ve}}\!\! \!d\bo{z} \, e^{i \bo{\xi^c} \cdot \bo{z}} \!\! \int \!\!d\bo{\xi} 
\, F_{\bo{n}} (\bo{\tau}+\ve v_0^{-1} \bo{z}, \bo{\xi};x) e^{-i \bo{\xi}\cdot \bo{z}}
\ee
Equation \eqref{pnil2} suggests that in order to have an asymptotic expansion of $I_l^{\ve}(t)$  it is sufficient 
to consider the Taylor expansion of $F_{\bo{n}} (\bo{\tau}+\ve v_0^{-1} \bo{z}, \bo{\xi};x)$ and to prove that
for each term we can extend the integration w.r.t. $\bs z $ to 

\be\label{omega0}
\Omega_{0} \equiv \{\bo{z} \in \erre^l \;|\; -\infty < z_1 <z_2 <\ldots <z_l < \infty \}
\ee

\n
paying only a small error.
\n
Let us rewrite $\lan \phi_n^\ve , I_l^\ve (t) \ran $ in a more convenient form. From (\ref{pnil2}) and (\ref{fnn}) we have 
\begin{multline}
\lan \phi_n^\ve , I_l^\ve (t;\ve x + R_0, \cdot) \ran \!
=\! \lf( \f{\ve}{i v_0} \ri)^{\!l} \!
\f{1}{\sqrt{\ve}} \, e^{i\varphi_{\ve} (x)}e^{\f{i}{\ve} n \tau}\!\!\!  
\sum_{n_1 \ldots n_l=0 }^{\infty} 
\int_{\Omega_\ve} \!\!\! d\bs{z} \, e^{i \boldsymbol{\xi}^c \cdot \boldsymbol{z}} \!\!
\int \!\! d\boldsymbol{\xi} \,  e^{-i \bsx \cdot \bsz} 
 \prod_{j=1}^l 
\widehat{ \phi_{n_{j+1}}\phi_{ n_j }} (\xi_j ) \widehat V (\xi_j )  \\
\cdot \, e^{\frac{i}{2} \tau \xi_j^2 } 
e^{i \tau \xi_{j-1}\sum_{l=j}^l \xi_l } e^{-i\xi_j x } \, 
\eta ( x -\bs{\tau} \!\cdot \!\boldsymbol{\xi} - \ve v_0^{-1}  \bs{z} \!\cdot \! \boldsymbol{\xi} )\, 
e^{\frac{i\ve }{2v_0} \sum_{j=1}^l ( z_j \xi_j^2 + 2 \xi_{j-1}\sum_{k=j}^l z_k \xi_k )}
\label{hako}
\end{multline}
We introduce the following matrix valued function
\begin{equation}
M(\boldsymbol{z}) =
\begin{pmatrix}
z_1 & z_2 & z_3& \dots & z_l \\
z_2 & z_2& z_3 &     &   z_l \\
z_3 & z_3 & z_3& &        \\

\vdots & & & \ddots & \\
z_l & z_l &z_l &\ldots&  z_l
\end{pmatrix} \label{ike}
\end{equation}
and we write
\be
 \sum_{j=1}^l \left( z_j \xi_j^2 + 2 \xi_{j-1}\sum_{k=j}^l z_k \xi_k \right)=
 \bs{\xi} \cdot M(\bs{z} )\bs{\xi} 
\ee
Furthermore we have
\be
\prod_{j=1}^l 
e^{\frac{i}{2} \tau \xi_j^2 } e^{i \tau \xi_{j-1}\sum_{l=j}^l \xi_l }
= e^{\frac{i}{2}  \bs{\xi} \cdot M(\bs{\tau} )\bs{\xi} } 
= e^{\frac{i}{2}\tau\lf( \sum_{j=1}^l \xi_{j} \ri)^2 }
\ee
The series over $n_1 \ldots n_l=0$ can be explicitly  computed and each of them reconstruct the propagator of the harmonic oscillator. 
In fact
\begin{align}
\sum_{n_1 \ldots n_l=0 }^{\infty} e^{i\bsx^c \cdot \bsz}  
&\prod_{j=1}^l \Big[   \widehat{ \phi_{n_{j+1}}\phi_{ n_j }} (\xi_j ) \widehat V (\xi_j )        \Big] \de_{n_1 0} \nonumber\\
		&= \sum_{n_1 \ldots n_l=0 }^{\infty} ( 2\pi)^{-j/2}\prod_{j=1}^l \Big[ \widehat{V}(\xi_j) \int dx_j \phi_{n_{j+1} } (x_j)\phi_{n_{j} } (x_j) 
				e^{-i\xi_j x_j}  	
					e^{\f{i}{v_0} (n_{j+1}- n_j)z_j} \Big] 	\de_{n_1 0} \nonumber\\
		&=( 2\pi)^{-j/2} e^{\f{i}{v_0}(n+1/2) z_l} \lf\lan \phi_n , e^{-i \xi_n \cdot}  U\lf(\f{z_l -z_{l-1}}{v_0} \ri)e^{-i \xi_{n-1} \cdot}  \cdots U\lf(\f{z_1 }{v_0} \ri)
				e^{-i \xi_1 \cdot}  \phi_0\ri\ran
			\prod_{j=1}^l   \widehat{V}(\xi_j) \nonumber\\
		&= ( 2\pi)^{-j/2}  e^{\f{i}{v_0}(n+1/2) z_l}  \lf\lan \phi_n , \zeta_{z} (\boldsymbol{\xi} )\ri\ran   \prod_{j=1}^l   \widehat{V}(\xi_j)  \label{gomi}
\end{align}
where in \eqref{gomi} we have introduced the notation 
\be
\zeta_{z} (\boldsymbol{\xi} ) \equiv
 e^{-i \xi_n \cdot}  U\lf(\f{z_l -z_{l-1}}{v_0} \ri)e^{-i \xi_{n-1} \cdot}  \cdots U\lf(\f{z_1 }{v_0} \ri)
				e^{-i \xi_1 \cdot}  \phi_0
\ee
In the end we arrive at
\begin{multline}
\lan \phi_n^\ve , I_l^\ve (t;\ve x + R_0, \cdot) \ran \!
=\! \lf( \f{\ve}{i v_0\sqrt{2\pi}} \ri)^{\!l} \!
\f{1}{\sqrt{\ve}} \, e^{i \varphi_{\ve}(x)}e^{\f{i}{\ve} n \tau}\! 
\int_{\Omega_{\ve}} \!\!\!d\bs{z} \, e^{\f{i}{v_0}n z_l} \!\! 
\int \!\! d\boldsymbol{\xi}  \,e^{-i \bsx \cdot \bsz}
e^{\frac{i}{2}\tau\lf( \sum_{j=1}^l \xi_{j} \ri)^2 } \!
\lan \phi_n , \zeta_{z} (\boldsymbol{\xi} )\ran \\
\cdot \prod_{j=1}^l 
 \widehat V (\xi_j )  \,  e^{-i\xi_j x } 
\eta ( x -\bs{\tau} \cdot \boldsymbol{\xi} - \ve v_0^{-1}  \bs{z} \cdot \boldsymbol{\xi} )
e^{\frac{i\ve }{2v_0} \bs{\xi} \cdot M(\bs{z} )\bs{\xi} }
\label{tora}
\end{multline}

\n
Our main result on the asymptotic behavior of $\Psi^{\pm}_{\ve}$ for $\ve \to 0$ in the stationary case 
is summarized in the following theorem.

\begin{teo}\label{t1}
Let us fix  $k \in \enne$, $t>\tau$ and assume that $||| V\, \lan \cdot \ran^2 |||_{k+2, 1} < \infty$.

\n
If $R_0<a$  then 

\ba\label{minato}
&&\Psi^{+}_{\ve} (t) =  e^{-i\f{t}{\ve^2} H_0^{\ve}} 
\lf(\f{1}{\sqrt{\ve}} \, e^{i \varphi_{\ve}}  \sum_{h=0}^{k} \ve^h \, \II^{+}_{h}  \ri)
+ \mathcal{S}^{+}_{\ve}(t)
\ea

\n
and if   $R_0>a$ then   
\ba\label{minatu}
&&\Psi^{-}_{\ve} (t) = 
e^{-i\f{t}{\ve^2} H_0^{\ve}} 
\lf(\f{1}{\sqrt{\ve}} \, e^{i \varphi_{\ve}}  
\sum_{h=0}^{k} \ve^h \, \II^{-}_{h} \right) + \mathcal{S}^{-}_{\ve}(t)
\ea
where the coefficient of the expansions $\II^{\pm}_{h}$ are given by\
\be
\II^{\pm}_{0}=\eta \phi_0 \hspace{3cm}
\label{honsha}
\II^{\pm}_{h}= \sum_{l=1}^h \II_l^{h,\pm} \quad h\geqslant 1
\ee
and
\begin{multline}\label{norimono}
\lan \phi_n , \II_l^{h,\pm} (\ve x + R_0, \cdot) \ran = \f{1}{h ! (i \sqrt{2\pi})^l (\pm v_0)^h}  
e^{\f{i}{\ve}n\tau}
\int_{\Omega_0} d\bs{z} e^{\pm\f{i}{v_0}(n+1/2) z_l}
\int d\boldsymbol{\xi}  \,e^{-i \bsx \cdot \bsz}
e^{\frac{i}{2}\tau\lf( \sum_{j=1}^l \xi_{j} \ri)^2 }
\lf\lan \phi_n , \zeta_{z} (\boldsymbol{\xi} )\ri\ran   \\
\cdot \prod_{j=1}^l 
 \widehat V (\xi_j )  \,  e^{-i\xi_j x } 
  \sum_{m=0}^{h-l} \eta^m ( x -\bs{\tau} \cdot \boldsymbol{\xi}  ) \lf(-   \bs{z} \cdot \boldsymbol{\xi} \ri)^m 
		\lf( \frac{i }{2}  \bs{\xi} \cdot M(\bs{z} )\bs{\xi}  \ri)^{h-l-m}
\end{multline}
The remainders $\mathcal{S}^{\pm}_{\ve}(t)$ 
satisfy
\ba
&&\|\mathcal{S}^{\pm}_{\ve} (t) \|<C_k \,\ve^{k+1}
\ea
where $C_k$ depends on $t$, $v_0$, $V$, $\eta$.
\end{teo}

\n
Theorem \ref{t1} shows that for $t>\tau$ we have again a free evolution of a modified initial state.

\n
We notice that the factor $\ve^{-1/2}$ in (\ref{minato}), (\ref{minatu}) takes into account 
the change of scale between $R$ and $x$ and therefore
$\ve^{-1/2}\|\II_l^{h,\pm}\|$ does not depend on $\ve$.


\section{Proofs}

\n
In this section we shall prove theorems \ref{t2} and \ref{t1}. 
For the sake of simplicity we shall drop $\ve$ as index. We are not interested in explicitly determining
the dependence on $k$ of $C_k$ and $D_k$ and therefore many combinatorial factors appearing
in the following formulas will be included in $c$. 
Let us start with the non stationary case.

\vs
\n
{\bf Proof of theorem 2.2}

\n
We shall give the proof only for $\Psi^+ (t)$ with $R_0 >a$, since the  case of $\Psi^- (t)$ with $R_0<a$ can be treated exactly in the same way. Let us consider the expansion \eqref{dui} with $k=0$ 
\be
 \Psi^+ (t) = e^{-i \f t \ve^2 H_0^\ve}\Psi_{0}^+ +e^{-i \f t \ve^2 H_0^\ve} I_1 (t) + {\mathcal R}_1 (t)
\ee
Taking into account \eqref{re}, we have
\be
\|\Psi^+(t) - e^{-i \f t \ve^2 H_0^\ve}\Psi_{0}^+ \|
\leqslant (1+ t\,  \|V\|_{L^{\infty}} ) \sup_{ 0\leqslant s \leqslant t} \| I_1(s) \|
\label{mokuteki}
\ee
From (\ref{mokuteki}) we see that the problem is  reduced to estimate $\|I_1(t)\|$. Formula (\ref{pnil}) for $l=1$ reads 
\begin{equation}
\lan \phi_n , I_1 (t;\ve x + R_0, \cdot)\ran \!=\! 
\f{-i}{\sqrt{\ve}} \, e^{i \varphi_{\ve}(x)} \mis
\! \int_0^t \!\!\! ds \!\! \int \!\!d\xi \, \widehat V(\xi)  \widehat{\phi_n \phi_0}(\xi) 
\eta(x-s\xi) e^{\f{i}{2}s\xi^2} e^{-ix\xi} 
e^{-\f{i}{\ve} v_0 \lf( \tau \xi - \f{n+1/2}{v_0} s + s\xi \ri) }
\end{equation}
Therefore we have
\begin{align}
 \| I_1 (t)\|^2  &=  \sum_{n=0}^{\infty} \int\!\! dx 
\lf|
\int_0^t \!\! ds\!\! \int \!\!d\xi \; \widehat V(\xi)  \widehat{\phi_n \phi_0}(\xi) 
 \,\eta(x-s\xi) e^{\f{i}{2}s\xi^2} e^{-ix\xi} 
e^{-\f{i}{\ve} v_0 \lf( \tau \xi - \f{n+1/2}{v_0} s + s\xi \ri) }
\ri|^2  \nonumber \\
	& = \sum_{n=0}^{\infty} \int \!\! dx  \int_0^t \!\!ds\, ds' \!\! \int \!\!d\xi\, d\xi' \;
	\widehat V(\xi)  \widehat{\phi_n \phi_0}(\xi) 
	\overline{\widehat V(\xi')  \widehat{\phi_n \phi_0}(\xi') } \,\eta(x-s\xi)\, \eta(x-s'\xi') 
	e^{\f{i}{2}s\xi^2} e^{-\f{i}{2}s'\xi^{2}}  \nonumber \\
	&\qquad \cdot\,   e^{-ix\xi} e^{ix\xi'}
	e^{-\f{i}{\ve} v_0 \lf( \tau \xi - \f{n+1/2}{v_0} s + s\xi \ri) }
	e^{\f{i}{\ve} v_0 \lf( \tau \xi' - \f{n+1/2}{v_0} s' + s'\xi' \ri) } \label{ichi}
\end{align}
The series over $n$ in \eqref{ichi} can be explicitly computed as follows.
\begin{align}
\sum_{n=0}^{\infty}  \widehat{\phi_n \phi_0}(\xi) \overline{ \widehat{\phi_n \phi_0}(\xi') }
	 e^{-i(n+1/2)\f{ (s'-s)}{\ve}   } 
	 & = \f{1}{2\pi} \sum_{n=0}^{\infty} 
	 \int dy \,dy'\; \phi_n(y) \phi_0 (y) e^{-i y \xi}  \phi_n(y') \phi_0 (y') e^{i y' \xi'} e^{-i\f{n (s'-s)}{\ve}   } \nonumber \\
	 & = \f{1}{2\pi} 
	 \lf(e^{-i  \xi'\cdot} \phi_0 , U\lf(  \f{ s'-s}{\ve}\ri)e^{-i  \xi\cdot} \phi_0\ri) \nonumber \\
	 &\equiv  \f{1}{2\pi}  W_{s,s'}(\xi, \xi') \label{go}
\end{align}
Substituting \eqref{go} into \eqref{ichi} we obtain
\begin{multline}
 \| I_1 (t)\|^2 = \! \int \!\!dx  \!\!\int_0^t \!\! ds\, ds' \!\!\int \!\!d\xi\, d\xi' \; \widehat{V}(\xi) \overline{\widehat{V}(\xi')}\,
	\eta(x-s\xi) \eta(x-s'\xi') e^{-ix\xi} e^{ix\xi'}e^{\f{i}{2}s\xi^2} e^{-\f{i}{2}s'\xi^{2}}  W_{s,s'}(\xi, \xi')  \\
	 \cdot \,
	e^{-\f{i}{\ve} v_0 \lf( \tau   + s \ri)\xi }
	e^{\f{i}{\ve} v_0 \lf( \tau   + s' \ri)\xi' } \label{ni}
\end{multline}
The estimate of \eqref{ni} can be easily obtained exploiting a non-stationary phase argument. We consider the identity
\begin{equation}
e^{-\f{i}{\ve} v_0 \lf( \tau   + s \ri)\xi } =
\lf( i\f{\ve}{v_0( \tau+s) } \ri)^{k+1} \partial_\xi^{k+1} 
e^{-\f{i}{\ve} v_0 \lf( \tau   + s \ri)\xi }
\end{equation}
and we integrate by parts $k+1$ times. Hence
\begin{align}
\| I_1 (t)\|^2 
		 &=\!\f{1}{2\pi}\!\lf( \!i\f{\ve}{v_0 } \!\ri)^{\!\!2k+2} \!\!\!\int \!\!dx \!\! \int_0^t \!\! ds\, ds'\frac{1}{( \tau+s)^{k+1}( \tau+s')^{k+1}} \!\int \!\! d\xi\, d\xi' \!   \sum_{ \substack{j_1,\ldots,j_5=0 \\ |\bs j|=k+1}}^{k+1}\!
		\binom{k+1}{\bs j} \!\!\sum_{ \substack{j_1',\ldots,j_5'=0 \\ |\bs j'|=k+1}}^{k+1}\!
		\binom{k+1}{\bs j'}  \nonumber \\ 
		& \quad\cdot \, \eta^{j_1}(x-s\xi) \, (-s)^{j_1} \widehat{V}^{j_2}(\xi) 
		 \partial_{\xi}^{j_3} e^{\f{i}{2}s\xi^2}  e^{-ix\xi} (-ix)^{j_4} \partial_{\xi}^{j_5} \partial_{\xi'}^{j_5'}W_{s,s'}(\xi, \xi') \nonumber \\
		& \quad\cdot \,
		  \eta^{j_1'}(x-s'\xi') \, (-s')^{j_1'} \widehat{V}^{j_2'}(\xi') 
		 \partial_{\xi}^{j_3}e^{\f{i}{2}s'\xi'^2}  e^{-ix\xi'} (-ix')^{j_4'} \label{san}
\end{align}
where 
\be
\dbinom{k}{\bs j} = \dfrac{k!}{j_1 ! \, j_2! \, j_3! \, j_4! \, j_5 !}
\ee
is the standard multinomial factor.
From (\ref{san}) we have
\begin{multline}
 \| I_1 (t)\|^2 
		  \leqslant c\lf( \!\f{\ve}{\tau v_0 }\! \ri)^{\!2k+2} \!\!\int \!\!dx\!\!  \int_0^t \!\! ds\, ds' \!\! \int \!\! d\xi\, d\xi' \! 
		\sum_{ \substack{j_1,\ldots,j_5=0 \\ |\bs j|=k+1}}^{k+1}\!
		\binom{k+1}{\bs j} \! 
		 \sum_{ \substack{j_1',\ldots,j_5'=0 \\ |\bs j'|=k+1}}^{k+1} \!
		\binom{k+1}{\bs j'} |\eta^{j_1}(x-s\xi)| \, s^{j_1}   \\
				\cdot \, |\widehat{V}^{j_2}(\xi) |
		  |\xi|^{j_3}
		  |x|^{j_4} 
		|\partial_{\xi}^{j_5} \partial_{\xi'}^{j_5'}W_{s,s'}(\xi, \xi')| 
		  |\eta^{j_1'}(x-s'\xi')| \, s'^{j_1'} |\widehat{V}^{j_2'}(\xi')| 
		  |s'\xi'|^{j_3'} |x'|^{j_4'} \label{yon}
\end{multline}
We interchange the integrals and we estimate the integration w.r.t. the variable $x$ using Cauchy-Schwartz inequality
\begin{equation}
\int\!\! dx \, |\eta^{j_1}(x-s\xi)| |x|^{j_4} |\eta^{j_1'}(x-s'\xi')||x'|^{j_4'} \leqslant c \, \|\eta^{j_1} \langle \cdot \rangle^{j_4} \|_{L^2} \langle s\xi \rangle^{j_4}
\|\eta^{j_1'} \langle \cdot \rangle^{j_4'} \|_{L^2} \langle s'\xi' \rangle^{j_4'}
\label{nana}
\end{equation}
Moreover  from the definition \eqref{go} and Cauchy-Schwartz inequality we have
\begin{multline}
|\partial_{\xi}^{j_5} \partial_{\xi'}^{j_5'}W_{s,s'}(\xi, \xi')|= 
\lf| \lf(e^{-i  \xi'\cdot}y^{j_5'} \phi_0 , U\lf(  \f{ s'-s}{\ve}\ri)e^{-i  \xi\cdot} y^{j_5}\phi_0\ri)\ri|
 \leqslant
\|y^{j_5} \phi_0\| \|y^{j_5'} \phi_0\| \\
=\f{1}{2\sqrt{\pi}}\sqrt{\Gamma\lf( j_5 +\f{1}{2} \ri) \Gamma\lf( j_5' +\f{1}{2} \ri) }
\label{roku}
\end{multline}
If we substitute \eqref{nana} and \eqref{roku} into \eqref{yon} we have
\begin{align}
 | I_1 (t)\| &\leqslant c 
		\lf( \!\f{\ve}{\tau v_0 \!} \ri)^{\!k+1}  \!\! \int_0^t \!\! ds \! \int \!\! d\xi  
		\sum_{ \substack{j_1,\ldots,j_5=0 \\ |\bs j|=k+1}}^{k+1}
		  \|\eta^{j_1} \langle x \rangle^{j_4} \|_{L^2} \langle s\xi \rangle^{j_4}  \, s^{j_1} |\widehat{V}^{j_2}(\xi) |
	 |s\xi|^{j_3}  \nonumber \\
		& \leqslant c\;\ve^{k+1}      \f{\lan t\ran^{k+2}}{(\tau v_0)^{k+1} }  
		\sum_{ \substack{j_1,\ldots,j_5=0 \\ |\bs j|=k+1}}^{k+1}
		\|   \eta^{j_1} \langle \cdot \rangle^{j_4} \|_{L^2} 
		\| \widehat{V}^{j_2} \langle \cdot \rangle^{j_3+j_4}\|_{L^1}  \nonumber \\
		&\leqslant c\;\ve^{k+1}      \f{\lan t\ran^{k+2}}{(\tau v_0)^{k+1} } 
		   ||| \eta |||_{k+1,2} ||| \widehat{V} |||_{k+1,1} \label{mugi}
\end{align}
The proof of the theorem now  follows from \eqref{mokuteki} and \eqref{mugi}.

\vs
\hfill $\Box$

\vs

\n
For the proof of theorem 2.3 it will be useful the following technical lemma.

\vs

\n
\begin{lem}
Let $\zeta_{t_1 \ldots t_{n-1}} : \erre^n \to L^2(\erre) $ be defined by
\begin{equation}
\zeta_{t_1 \ldots t_{n-1}} (\boldsymbol{\xi})  \equiv
e^{-i \xi_n \cdot}  U(t_{n-1})e^{-i \xi_{n-1} \cdot}  \cdots U(t_1) e^{-i \xi_1 \cdot}  \phi_0 \label{futari}
\end{equation}
Then $\zeta_{t_1 \ldots t_{n-1}} \in C^{\infty} (\erre^n ; L^2(\erre))$ and for every multi-index $\bs \al  =(\al_1, \ldots ,\al_n) \in \enne^n$ there
exists $c_{\bs \al}$, independent of  $t_i$ for $i=1,\ldots , n-1$, such that 
the following estimate holds 
\begin{equation}
 \|\partial_{\bs \xi}^{\bs \al} \zeta_{t_1 \ldots t_{n-1}} (\bs \xi) \|_{L^2} \equiv \lf\| \left(  \prod_{j=1}^n \partial_{\xi_j}^{\al_j} \!\!\right)  \zeta_{t_1 \ldots t_{n-1}}(\bs \xi) \ri\|_{L^2}
 \leqslant  c_{\bs \al} \lf( \sum_{h=1}^{n} \lan \xi_h \ran \ri)^{\!\!|\bs \al|}
\label{sato}
\end{equation}
\end{lem}
\begin{dem}
The derivative  $\partial_{\bs \xi}^{\bs \al} \zeta_{t_1 \ldots t_{n-1}} (\bs \xi)$ is explicitly given by
\begin{equation}
\partial_{\bs \xi}^{\bs \al} \zeta_{t_1 \ldots t_{n-1}} (\bs \xi) = (-i)^{|\bs \al| }
e^{-i \xi_n \cdot} Q^{\al_n} U(t_{n-1})e^{-i \xi_{n-1}  \cdot} Q^{\al_{n-1}}  \cdots U(t_1) e^{-i \xi_1 \cdot}   Q^{\al_{1}}\phi_0
\label{neko}
\end{equation}
where $Q$ denotes the multiplication operator by the independent variable $y$. 
In order to estimate the norm of \eqref{neko} we shall move all the $Q$'s to the right until they act on $\phi_0$ using the following identities
\begin{equation}
\begin{cases}
Q \, U(t) = U(t)  
\lf( Q \cos t  - D  \sin t  \ri) \\
D\, U(t) = U(t) 
\lf(  Q \sin t + D  \cos t  \ri)
\end{cases}
\label{inu}
\end{equation}
where  $D = i \f{d}{dy}$.  
Let us first see how the strategy works for the case $|\bs \al|=1$.
\begin{align}
&\partial_{\xi_j} \zeta_{t_1 \ldots t_{n-1}}(\bs \xi) 
  = (-i) e^{-i \xi_n \cdot}   \cdots  e^{-i \xi_j \cdot}  QU(t_{j-1})  \cdots U(t_1) e^{-i \xi_1 \cdot}   \phi_0 \nonumber \\
&\quad =(-i) e^{-i \xi_n \cdot}  \cdots  U(t_{j-1}) \lf( Q \cos t_{j-1}  - D \sin t_{j-1} \ri)
				e^{-i \xi_{j-1} \cdot}  U(t_{j-2})\cdots U(t_1) e^{-i \xi_{j-1} \cdot}   \phi_0 \nonumber \\
& \quad=(-i) e^{-i \xi_n \cdot}  \cdots U(t_{ j-1})	e^{-i \xi_{j-1} \cdot} 
				 \lf( Q \cos t_{j-1}  - D \sin t_{j-1}  - \xi_{j-1} \sin t_{j-1} \ri)
			 	U(t_{j-2})\cdots U(t_1) e^{-i \xi_1 \cdot}   \phi_0 \nonumber  \\
\end{align}		 	
Iterating the procedure we arrive at 
\begin{equation}
\partial_{\xi_j} \zeta_{t_1 \ldots t_{n-1}}(\bs \xi)	 
	=\! ( -i) e^{-i \xi_n \cdot}   \cdots U(t_1) e^{-i \xi_1 \cdot} 
			 	\Big(\! Q \cos \Big( \!\sum_{m=1}^{j-1} t_m \!\Big)  - D \sin \Big(\!\sum_{m=1}^{j-1} t_m \! \Big)  -\sum_{m=1}^{j-1} \xi_m \sin\Big(\!\sum_{h=m}^{j-1} t_h\! \Big)\! \Big)\phi_0
\label{tanuchi}
\end{equation}
From equation \eqref{tanuchi} we obtain the estimate
\[
\|   \partial_{\xi_j} \zeta_{t_1 \ldots t_{n-1}}(\bs \xi)  \|_{L^2}
\leqslant \|Q \phi_0\|_{L^2} + \|D\phi_0\|_{L^2} +\! \sum_{m=1}^{j-1} |\xi_m| 
 \leqslant \! \lf( \! 1+ \lf\| Q \phi_0 \ri\|_{L^2} + \lf\| D \phi_0 \ri\|_{L^2} \ri)\!
\lf( \sum_{m=1}^{n} \lan \xi_m \ran \!\ri)
\]
Derivatives of any order can be estimated by iteration of the procedure used in \eqref{tanuchi}, 
starting from the most left derivative and going to the right. This leads to
\begin{equation}
\|\partial_{\bs \xi}^{\bs \al} \zeta_{t_1\ldots t_{n-1}} (\bs \xi) \|_{L^2} 
\leqslant
\! \lf\| \prod_{j=1}^n
\lf[ \! \lf(\! Q \cos \lf(\! \sum_{m=1}^{j-1} t_m \! \ri) \! - D \sin \lf(\! \sum_{m=1}^{j-1} t_m \! \ri)\! -\sum_{m=1}^{j-1} \xi_m \sin\lf(\! \sum_{h=m}^{j-1} t_h \! \ri) \!  \ri) \! \ri]^{\al_j}\!\!\!\!\!
\phi_0
\ri\|_{L^2} \label{katsudon}
\end{equation}
From  estimate \eqref{katsudon} it is clear that there exists $c_{\bs \al}$,  depending on
suitable norms of $\phi_0$, such that (\ref{sato}) holds.
\end{dem}

\n
In the next proposition  we derive an expansion for the quantity $\lan \phi_n , I_l (t) \ran$ which is the key ingredient for the proof of theorem 2.3.

\begin{prop}
\label{tatemono}
\n
Let us consider $\Psi^+_{\ve, 0}$ as initial state and
let us fix $R_0 <a$,  $t>\tau$ and $k , l \in \enne$, with $l \leqslant k$.  Then
there  exists a constant $C$,  
depending on $k,l,v_0,\tau,t,\eta, V$,  such that
\begin{equation}\label{expi}
\lan \phi_n , I_l (t) \ran = \sum_{h=l}^k \ve^h \lan \phi_n ,  \II_l^{h,+}  \ran 
+ \ve^{k+1} \lan \phi_n , {\mathcal T} (t)\ran
\end{equation}
where
\begin{multline}
\lan \phi_n ,  \II_l^{h,+} (\ve x + R_0,\cdot) \ran = \f{1}{h ! (i \sqrt{2\pi})^l v_0^h}  
\f{1}{\sqrt{\ve}} \, e^{i\varphi_\ve(x)}\mis \!
\int_{\Omega_0} \!\!\! d\bs{z}\,  e^{\f{i}{v_0}(n+1/2) z_l}\!\!
\int d\boldsymbol{\xi}  \,e^{-i \bsx \cdot \bsz}
e^{\frac{i}{2}\tau\lf( \sum_{j=1}^l \xi_{j} \ri)^2 }
\lf\lan \phi_n , \zeta_{z} (\boldsymbol{\xi} )\ri\ran   \\
\cdot \prod_{j=1}^l 
 \widehat V (\xi_j )  \,  e^{-i\xi_j x } 
  \sum_{m=0}^{h-l} \eta^m ( x -\bs{\tau} \!\cdot \!\boldsymbol{\xi}  ) \lf(-   \bs{z} \!\cdot \!\boldsymbol{\xi} \ri)^m 
		\lf( \frac{i }{2}  \bs{\xi} \cdot M(\bs{z} )\bs{\xi}  \ri)^{\!\!h-l-m}
\end{multline}
and
\be
\| {\mathcal T} (t) \| \leqslant C \ve^{k+1}
\ee
\end{prop}
\begin{dem}
As a first step  we consider the Taylor expansion of the $\ve$-dependent part of the integrand in \eqref{tora} up to order $k-l$. We have

\n
\ba
&&\eta ( x -\bs{\tau}\! \cdot \!\boldsymbol{\xi} - \ve v_0^{-1}  \bs{z} \!\cdot\! \boldsymbol{\xi} ) \, e^{\frac{i\ve }{2v_0} \bs{\xi} \cdot M(\bs{z} )\bs{\xi} }
	 = \sum_{j=0}^{k-l} \f{1}{j !} \lf( \f \ve v_0 \ri)^j \sum_{m=0}^j \eta^m ( x -\bs{\tau} \!\cdot \!\boldsymbol{\xi}  ) \lf(-   \bs{z} \!\cdot \!\boldsymbol{\xi} \ri)^m 		\lf( \frac{i }{2}\; \bs{\xi} \cdot M(\bs{z} )\bs{\xi}  \ri)^{j-m} \nonumber \\
	&& +\f{1}{(k-l+1) !} \lf( \f \ve v_0 \ri)^{\! k-l+1} \!\!\int_0^1 \!\!\! d\theta
		\sum_{m=0}^{k-l+1}  \eta^m ( x -\bs{\tau} \!\cdot \!\boldsymbol{\xi}  - \ve\theta v_0^{-1}  \bs{z} \!\cdot \!\boldsymbol{\xi}) 
	\lf(-   \bs{z} \!\cdot \! \boldsymbol{\xi} \ri)^m  
	\lf( \frac{i }{2} \; \bs{\xi} \!\cdot \!M(\bs{z} )\bs{\xi}  \ri)^{\!k-l+1-m}\! \nonumber\\
&&\cdot \, e^{\frac{i\ve\theta }{2v_0}  \bs{\xi} \cdot M(\bs{z} )\bs{\xi} } (1-\theta)^{k-l+1}
\label{rengagio}
\ea
The corresponding expansion of $\lan \phi_n , I_l (t) \ran$ is given by
\begin{equation}
\lan \phi_n , I_l (t) \ran = \sum_{h=l}^k \ve^h \lan \phi_n , \tilde I_l^h (t) \ran 
+ \ve^{k+1}  \lan \phi_n, \tilde{{\mathcal T}} (t) \ran
\label{senro}
\end{equation}
with
\begin{multline}
\lan \phi_n , \tilde I_l^h (t;\ve x + R_0, \cdot) \ran = \f{1}{h ! (i \sqrt{2\pi})^l v_0^h}  
\f{1}{\sqrt{\ve}} \, e^{i \varphi_\ve (x)} \mis\!\!
\int_{\Omega_\ve} \!\!\!d\bs{z}\,  e^{\f{i}{v_0}(n_1/2) z_l} \!\!\!
\int \!\! d\boldsymbol{\xi}  \,e^{-i \bsx \cdot \bsz}
e^{\frac{i}{2}\tau\lf( \sum_{j=1}^l \xi_{j} \ri)^2 }\!
\lf\lan \phi_n , \zeta_{z} (\boldsymbol{\xi} )\ri\ran   \\
 \cdot \prod_{j=1}^l 
 \widehat V (\xi_j )  \,  e^{-i\xi_j x } 
  \sum_{m=0}^{h-l} \eta^m ( x -\bs{\tau} \!\cdot \!\boldsymbol{\xi}  ) \lf(-   \bs{z} \!\cdot \!\boldsymbol{\xi} \ri)^m 
		\lf( \frac{i }{2}  \bs{\xi} \!\cdot \! M(\bs{z} )\bs{\xi}  \ri)^{h-l-m}
\end{multline}
and
\[
\lan \phi_n, \tilde{{\mathcal T}} (t) \ran=\sum_{m=0}^{k-l+1}  \tilde{{\mathcal T}}_m^n (t)
\]
\begin{multline}
\tilde{{\mathcal T}}_m^n (t;\ve x + R_0) =   \f{e^{i \varphi_\ve (x)} \mis}{(k-l+1) !v_0^{ k-l+1}}
\int_{\Omega_\ve}\!\!\!  d\bs{z} \,  e^{\f{i}{v_0}n z_l}\!\!\! 
\int \!\!  d\boldsymbol{\xi}  \,e^{-i \bsx \cdot \bsz}
e^{\frac{i}{2}\tau\lf( \sum_{j=1}^l \xi_{j} \ri)^2 }
\lf\lan \phi_n , \zeta_{z} (\boldsymbol{\xi} )\ri\ran  
\prod_{j=1}^l 
 \widehat V (\xi_j )  \,  e^{-i\xi_j x } \\
 \cdot\int_0^1 \!\!\! d\theta \, 
 \eta^m ( x -\bs{\tau} \!\cdot \!\boldsymbol{\xi}  - \ve\theta v_0^{-1}  \bs{z}\! \cdot \! \boldsymbol{\xi}) \lf(-   \bs{z} \!\cdot \!\boldsymbol{\xi} \ri)^m 
 \lf( \frac{i }{2}\; \bs{\xi} \!\cdot \!M(\bs{z} )\bs{\xi}  \ri)^{\!k-l+1-m} \!\!
 e^{\frac{i\ve\theta }{2v_0} \,\bs{\xi} \cdot M(\bs{z} )\bs{\xi} } (1-\theta)^{k-l+1}
\end{multline}
We have  to prove that each term in the r.h.s. of \eqref{senro} is well  defined, i.e. $\tilde{I}_l^h (t)$ and $\tilde{\mathcal T}(t)$ must be  elements  of  $ L^2 (\erre^2)$.  We  show this fact  only for the most singular term  $\tilde{\mathcal T}(t)$, since the other terms can be treated in a similar way. 



\n
The square of the norm of $\tilde{{\mathcal T}}_m^n(t)$ is given by
\begin{align}
\sum_{n=0}^{\infty} \! \| \tilde{{\mathcal T}}_m^n (t)\|_{L^2}^2 
&=c \!\!\int \!\! dx \!\!\int_{\Omega_\ve} \!\!\! d\bs{z} \, d\bs{z}' \!\!\int \!\!d\boldsymbol{\xi} \,  d\boldsymbol{\xi}'\,
e^{-i \bsx \cdot \bsz}e^{i \bsx' \cdot \bsz'}
e^{\frac{i}{2}\tau\lf( \sum_{j=1}^l \xi_{j} \ri)^2 }
e^{-\frac{i}{2}\tau\lf( \sum_{j=1}^l \xi_{j} ' \ri)^2 }
\lf\lan \!\! \zeta_{ z'} (\boldsymbol{\xi}'  )  ,     U \lf(   \f{z_l '-z_l}{v_0} \ri)  \zeta_{z} (\boldsymbol{\xi} )\!\!\ri\ran  \nonumber \\
& \cdot \prod_{j=1}^l  \widehat V (\xi_j ) \overline{\widehat V }(\xi_j ') 
\,  e^{-i\xi_j x } e^{i\xi_j' x } 
\int_0^1 d\theta\,
 \eta^m ( x -\bs{\tau} \cdot \boldsymbol{\xi}  - \ve\theta v_0^{-1}  \bs{z} \cdot \boldsymbol{\xi}) \lf(-   \bs{z} \cdot \boldsymbol{\xi} \ri)^m \nonumber \\
& \cdot \lf( \frac{i }{2}\; \bs{\xi} \cdot M(\bs{z} )\bs{\xi}  \ri)^{k-l+1-m}
e^{\frac{i\ve\theta }{2v_0} \, \bs{\xi} \cdot M(\bs{z} )\bs{\xi} } (1-\theta)^{k-l+1} 
 \int_0^1 d\theta'
 \eta^{m} ( x -\bs{\tau} \cdot \boldsymbol{\xi}'  - \ve\theta' v_0^{-1}  \bs{z}' \cdot \boldsymbol{\xi}') 
  \nonumber \\
&\cdot  \lf(-   \bs{z}' \cdot \boldsymbol{\xi}' \ri)^{m} \lf(- \frac{i }{2} \;\bs{\xi}' \cdot M(\bs{z}' )\bs{\xi}'  \ri)^{k-l+1-m}
e^{-\frac{i\ve\theta' }{2v_0} \, \bs{\xi}' \cdot M(\bs{z}' )\bs{\xi}' } (1-\theta')^{k-l+1}
\end{align}
where, once again, we have reconstructed the propagator of the harmonic oscillator exploiting the formula
\ba
&&\lf\lan  \zeta_{ z'} (\boldsymbol{\xi}'  )  ,     U \lf(   \f{z_l '-z_l}{v_0} \ri)  \zeta_{z} (\boldsymbol{\xi} )\ri\ran = \sum_{n=0}^{\infty} e^{\f{i}{v_0} (n+1/2) (z_l - z'_l)} \lan \phi_n , \zeta_z (\bo{\xi})\ran \overline{\lan \phi_n , \zeta_{z'} (\bo{\xi'})\ran }
\ea

\n
In order to control the convergence of the integrals over $\Omega_\ve$,  
we integrate by parts $k+2$ times w.r.t. to $\xi_j$ and to $\xi_j'$ respectively, for $j=1,\ldots , l$. For the sake of simplicity, we explicitly 
perform the computation in the case $l=k=2$, the general case being more delicate only from the notational point
of view. Then we are reduced to estimate the following quantity for $m=0,1$
\begin{align}
\label{Gzz}
\sum_{n=0}^{\infty}  \| \tilde{{\mathcal T}}_m^n \|_{L^2}^2
&= c\int \!\! dx \!\! \int_{\Omega_\ve} \!\! d\bs{z} d\bs{z}' \!\! \int \!\! d\boldsymbol{\xi}  d\boldsymbol{\xi}'\,
e^{-i \bsx \cdot \bsz}e^{i \bsx' \cdot \bsz'}
e^{\frac{i}{2}\tau\lf( \xi_{1}+\xi_2 \ri)^2 }
e^{-\frac{i}{2}\tau\lf(  \xi_{1}' +\xi_2' \ri)^2 }
\lf\lan \! \zeta_{ z'} (\boldsymbol{\xi}'  ),U \lf(   \f{z_2 '-z_2}{v_0} \ri)  \zeta_{z} (\boldsymbol{\xi} )\!\ri\ran \nonumber \\
& \cdot\, \widehat V (\xi_1 ) \widehat V (\xi_2 ) \overline{\widehat V }(\xi_1 ') \overline{\widehat V }(\xi_2 ') 
\,  e^{-i\lf( \xi_{1}+\xi_2 \ri)x } e^{i (\xi_{1}' +\xi_2' )x } 
\int_0^1 d\theta \,
 \eta^m ( x -\bs{\tau} \cdot \boldsymbol{\xi}  - \ve\theta v_0^{-1}  \bs{z} \cdot \boldsymbol{\xi}) \lf(-   \bs{z} \cdot \boldsymbol{\xi} \ri)^m \nonumber \\
& \cdot \, \lf( \frac{i }{2} \; \bs{\xi} \cdot M(\bs{z} )\bs{\xi}  \ri)^{1-m}
e^{\frac{i\ve\theta }{2v_0} \bs{\xi} \cdot M(\bs{z} )\bs{\xi} } (1-\theta)
 \int_0^1 d\theta' \, 
 \eta^{m} ( x -\bs{\tau} \cdot \boldsymbol{\xi}'  - \ve\theta' v_0^{-1}  \bs{z}' \cdot \boldsymbol{\xi}') \lf(-   \bs{z}' \cdot \boldsymbol{\xi}' \ri)^{m} \nonumber \\
& \cdot\, \lf(- \frac{i }{2} \; \bs{\xi}' \cdot M(\bs{z}' )\bs{\xi}'  \ri)^{1-m}
e^{-\frac{i\ve\theta' }{2v_0} \, \bs{\xi}' \cdot M(\bs{z}' )\bs{\xi}' } (1-\theta')\nonumber \\
&\equiv \int_{\Omega_\ve} \!\! d\bs{z} \, d\bs{z}' \, G( \bs{z}, \bs{z}')
\end{align}
We shall derive a pointwise estimate of $G$. Let us introduce the shorthand  notation $\partial_1^{j_i} = \partial_{\xi_1}^{j_i}$ and $\partial_1^{j_i'} = \partial_{\xi_1'}^{j_i'}$ and let us consider  the identity
\be
(-iz_1)^4 e^{-i \bsx \cdot \bsz} = \partial_1^4 e^{-i \bsx \cdot \bsz}
\ee
Then integrating by parts four times w.r.t. $\xi_1$  and $\xi_1 '$  in (\ref{Gzz}) we have
\begin{align}
z_1^4  z_1 '^4 G( \bs{z}, \bs{z}')
&= \int \!\! dx
 \!\! \int \!\! d\boldsymbol{\xi} \!\! \int \!\! d\boldsymbol{\xi}'\,
 e^{-i \bsx \cdot \bsz}e^{i \bsx' \cdot \bsz'}
 \widehat V (\xi_2 ) \overline{\widehat V }(\xi_2 ') 
\,  e^{-i \xi_2 x } e^{i  \xi_2' x } 
\sum_{ \substack{j_1,\ldots,j_7=0 \\ |\bs j|=4}}^4 
\binom{4}{\bs j}
\sum_{ \substack{j_1',\ldots,j_7'=0 \\ |\bs j'|=4}}^4 
\binom{4}{\bs j'} \nonumber \\
& \cdot \,
\partial_1^{j_1} e^{\frac{i}{2}\tau\lf( \xi_{1}+\xi_2 \ri)^2 }
\partial_1^{j_2}\partial_1^{j_2'}
\lf\lan\zeta_{ z'} (\boldsymbol{\xi}'),U \lf(\f{z_2 '-z_2}{v_0} \ri)\zeta_{z}(\boldsymbol{\xi} )\ri\ran
\partial_1^{j_3}\widehat V (\xi_1 )
\partial_1^{j_4}e^{-i \xi_1 x } \nonumber \\
&\cdot \int_0^1 \!\! d\theta \, (1-\theta)\,
 \partial_1^{j_5}\eta^m ( x -\bs{\tau} \!\cdot \!\boldsymbol{\xi}  - \ve\theta v_0^{-1}  \bs{z}\! \cdot \! \boldsymbol{\xi})\,  \partial_1^{j_6}\lf\{ \lf(-   \bs{z} \cdot \boldsymbol{\xi} \ri)^m 
 \lf( \frac{i }{2}\; \bs{\xi} \cdot M(\bs{z} )\bs{\xi}  \ri)^{1-m} \ri\}
 \nonumber \\
&\cdot\,  \partial_1^{j_7}e^{\frac{i\ve\theta }{2v_0} \, \bs{\xi} \cdot M(\bs{z} )\bs{\xi} } \, 
\partial_1^{j_1'} e^{-\frac{i}{2}\tau\lf( \xi_{1}'+\xi_2' \ri)^2 }
\partial_1^{j_3'}\overline{\widehat V }(\xi_1' )
\, \partial_1^{j_4'}e^{i \xi_1 x } \!\!
\int_0^1 \!\! d\theta' (1-\theta')
 \partial_1^{j_5'}\eta^m ( x -\!\bs{\tau} \!\cdot \!\boldsymbol{\xi}'  - \!\ve\theta' v_0^{-1}  \bs{z}' \! \cdot \! \boldsymbol{\xi}') \nonumber \\
&\cdot \, \partial_1^{j_6}\lf\{ \lf(-   \bs{z}' \cdot \boldsymbol{\xi}' \ri)^m 
 \lf( -\frac{i }{2}\; \lan\bs{\xi} '\cdot M(\bs{z}' )\bs{\xi}'  \ri)^{1-m} \ri\}
\partial_1^{j_7'}e^{-\frac{i\ve\theta '}{2v_0} \, \bs{\xi}' \cdot M(\bs{z}' )\bs{\xi}' } 
\label{gomibako}
\end{align}

\n
Using  \eqref{sato} and exploiting the fact that $\|  \ve v_0^{-1}  \bs{z}\|\leqslant |t| $ for $\bs{z} \in \Omega_{\ve}$, we have
\begin{align}
z_1^4 z_1'^4  |G( \bs{z}, \bs{z}')|
 &\leqslant c |\bsz||\bsz'|\int \!\! dx\int_0^1 \!\! d\theta \, d\theta' \, (1-\theta)(1-\theta')\int \!\! d\boldsymbol{\xi}  d\boldsymbol{\xi}'\,
|\widehat V (\xi_2 )| |\widehat V (\xi_2 ')| \nonumber\\
&\cdot \sum_{ \substack{j_1,\ldots,j_7=0 \\ |\bs j|=4}}^4 
\sum_{ \substack{j_1',\ldots,j_7'=0 \\ |\bs j'|=4}}^4 \lan t\ran^{j_1+ j_5+j_1'+ j_5'} 
\lan \bsx \ran^{2+ j_1+j_2+j_7}|\widehat V^{j_3} (\xi_1)| 
 \lan x\ran^{j_4} 
 |\eta^{m+j_5} ( x -\bs{\tau} \!\cdot \!\boldsymbol{\xi}  - \ve\theta v_0^{-1}  \bs{z} \!\cdot \!\boldsymbol{\xi})| \nonumber\\
&\cdot \, \lan \bsx' \ran^{2+ j_1'+j_2'+j_7'}
|\widehat V^{j_3} (\xi_1')| \lan x\ran^{j_4'}
   |\eta^{m+j_5} ( x -\bs{\tau} \!\cdot \!\boldsymbol{\xi}'  - \ve\theta v_0^{-1}  \bs{z}' \!\cdot \!\boldsymbol{\xi}')|
  \label{fugu}
\end{align}
We use Cauchy-Schwartz inequality to estimate the integral w.r.t. $x$ and we obtain

\begin{align}
 z_1^4 z_1'^4  |G( \bs{z}, \bs{z}')| 
 &\leqslant c \, \lan t\ran^{8} |\bsz||\bsz'|\lf(
\sum_{ \substack{j_1,j_2,j_3,j_4=0 \\ j_1+j_2+j_3+j_4=4}}^4 \!\! 
\|\eta^{m+j_4}\lan \cdot \ran^{j_2}\|_{L^2} \!
\int \!\!d\xi_1 \,d\xi_2 \, |\widehat V (\xi_2)|
\lan \bsx \ran^{2+ j_1+j_2}
|\widehat V^{j_3} (\xi_1)| \ri)^{\!\!2}\nonumber \\
&\leqslant c \, \lan t\ran^{8} |\bsz||\bsz'|
\lf(
\sum_{ \substack{j_1,j_2=0 \\ j_1+j_2=4}}^4 
\|\eta^{m+j_1}\lan \cdot \ran^{j_2}\|_{L^2}
\ri)^{\!\!2}
\lf(
\int \!\! d\xi 
\sum_{ \substack{j_1,j_2=,0 \\ j_1+j_2=4}}^4 
\lan \bsx \ran^{2+ j_1}|\widehat V^{j_2} (\xi)| 
\ri)^{\!\!4} \nonumber \\
&\leqslant c \,  | \bsz | | \bsz' | \lan t\ran^8 
 ||| \eta^m |||_{4,2}^2 |||\widehat V\lan\cdot \ran^2|||_{4,1}^4
\label{korosu}
\end{align}
Since the variables $z_1, z_1'$ play no  special role, we can repeat the same computation 
for each variable and we still obtain the same estimate as \eqref{korosu}. Summing up we arrive at
\be
|G( \bs{z}, \bs{z}')| \leqslant c \,  \lan \bsz \ran^{-3} \lan \bsz' \ran^{-3} \lan t\ran^8 
 ||| \eta^m |||_{4,2}^2 |||\widehat V\lan\cdot \ran^2|||_{4,1}^4
\label{umi}
\ee
The above estimate guarantees that (see (\ref{Gzz}) 
\be
\sum_{m=0,1} \! \left(\sum_{n=0}^{\infty} \| \tilde{\mathcal T}^{n}_{m} (t) \|^2_{L^2} \! \right)^{\!1/2} \!\!\! \leqslant 
c \,
\lan t\ran^4 \, |||\widehat V\lan\cdot \ran^2|||_{4,1}^2
 \sum_{m=0,1} ||| \eta^m |||_{4,2} 
\ee
\n
in the particular case $l=k=2$. In the general case, exploiting  the same argument, one obtain
\be
|G( \bs{z}, \bs{z}')| \leqslant c \, \lan \bsz \ran^{-(l+1)} \lan \bsz' \ran^{-(l+1)} \lan t\ran^{2(k+2)} \,
 ||| \eta^m |||_{k+2,2}^2 \, |||\widehat V\lan\cdot \ran^2|||_{k+2,1}^{2l}
\ee
and therefore
\be
\|\tilde{\mathcal T} (t)\|  \leqslant \sum_{m=0}^{k-l+1} \! \left(\sum_{n=0}^{\infty} \| \tilde{\mathcal T}^{n}_{m} (t) \|^2_{L^2} \!\right)^{\!1/2} \!\!\! \leqslant c \, \lan t\ran^{k+2} 
|||\widehat V\lan\cdot \ran^2|||_{k+2,1}^{l}\!\!
\sum_{m=0}^{k-l+1} ||| \eta^m |||_{k+2,2}
\ee
Using the same kind of arguments, it is easily shown that also  $\tilde{I}_l^h (t)$ belongs to $L^2(\erre^2)$ and this means that equation (\ref{senro}) is well defined. To conclude the proof of the proposition it remains to estimate the difference $\II_l^{h,+} - \tilde{I}_l^h (t)$. We have
\begin{multline}
\lan \phi_n ,  \II_l^{h,+}(\ve x +R_0,\cdot) - \tilde I_l^h(t;\ve x + R_0,\cdot) \ran 
= \f{e^{i \varphi_{\ve}(x)} \mis}{h ! (i \sqrt{2\pi})^l v_0^h}  
\f{1}{\sqrt{\ve}} \, 
\int_{\Omega_0\setminus \Omega_{\ve}} d\bs{z} e^{\f{i}{v_0}n z_l}
\int d\boldsymbol{\xi}  \,e^{-i \bsx \cdot \bsz}
e^{\frac{i}{2}\tau\lf( \sum_{j=1}^l \xi_{j} \ri)^2 } \\
\cdot \lf\lan \phi_n , \zeta_{z} (\boldsymbol{\xi} )\ri\ran 
\cdot \prod_{j=1}^l 
 \widehat V (\xi_j )  \,  e^{-i\xi_j x } 
  \sum_{m=0}^{h-l} \eta^m ( x -\bs{\tau} \cdot \boldsymbol{\xi}  ) \lf(-   \bs{z} \cdot \boldsymbol{\xi} \ri)^m 
		\lf( \frac{i }{2}  \bs{\xi} \cdot M(\bs{z} )\bs{\xi}  \ri)^{h-l-m}
\end{multline}
If we repeat once again the integration by parts procedure outlined above we 
obtain
\be
\| \II_l^{h,+} - \tilde I_l^h (t) \| \leqslant c \, 
\lan t\ran^{k+2} 
|||\widehat V\lan\cdot \ran^2|||_{k+2,1}^{l}
\sum_{m=0}^{k-l+1} ||| \eta^m |||_{k+2,2}
\int_{\Omega_0 \setminus \Omega_{\ve}} d\bsz \f{1}{\lan \bsz \ran^{k+2-h+l}}
\ee
We observe that
\be
\int_{\Omega_0 \setminus \Omega_{\ve}} d\bsz \f{1}{\lan \bsz \ran^{k+2-h+l}}
\leqslant 
\int_{\|\bsz\|\leqslant \ve^{-1}\lan t\ran} d\bsz \f{1}{\lan \bsz \ran^{k+2-h+l}}
\leqslant c\lf( \f{\ve}{\lan t\ran} \ri)^{k+1-h}
\ee
Therefore the expansion  \eqref{expi} is proved with

\be
\mathcal T (t) = \tilde{\mathcal T}(t) + \sum_{h=l}^k \ve^{h-k-1} (\II_l^{h,+} - \tilde I_l^h (t))
\ee
\end{dem}

\n
We are now in position to prove theorem 2.3.
\vs

\n
{\bf Proof of theorem 2.3}

\n
We give the detail for asymptotic expansion associated to $\Psi^+_{\ve,0}$ with $R_0<a$. The case 
with $\Psi^-_{\ve,0}$ with $R_0>a$ is completely analogous.
We consider formula  \eqref{dui} and we expand each $I_l(t)$, for  $l=1,\ldots ,k$,  according to proposition \ref{tatemono}.
Thus we obtain \eqref{minato} \eqref{honsha} and \eqref{norimono}. It remains to prove that the rest is  of
order $\ve^{k+1}$. From (\ref{re}) we have 
\be
\|{\mathcal R}_{k+1} (t) \| \leqslant \|V\|_{L^{\infty}} \int_0^t ds \, \| I_{k+1} (s) \|
\ee
It is straightforward to notice that, under the assumptions of theorem \ref{t2}, one has the estimate $\| I_{k+1} (s) \|\leqslant c \, \ve^{k+1}$ (we just
have to proceed as in proposition \ref{tatemono} integrating by parts $k+2$ times without the Taylor expansion) and therefore the proof is concluded.

\vs
\hfill $\Box$

\section{An application}
\n
Here we give an explicit computation of  the first and second order terms of the asymptotic expansion of $\Psi^+_{\ve}(t)$ in the stationary case and  discuss an application. From theorem 2.3 we have 
\be
\Psi^{+}_{\ve} (t) = e^{-i\f{t}{\ve^2} H_0^{\ve}}  
\lf[
\f{1}{\sqrt{\ve}} \, e^{i \varphi_\ve}  
\lf(
\II_0^{+}+\ve\, \II^{1,+}_1 + \ve^2 \lf( \II^{2,+}_1 + \II^{2,+}_2 \ri) \ri) \ri]
+ {\mathcal O}(\ve^3)
\ee
The first order term is given by
\be
\lan \phi_n , \II^{1,+}_1 ( \ve x+ R_0, \cdot ) \ran = 
\f{1}{i v_0 \sqrt{2\pi} } \mis \int_{\erre} dz \, e^{i \f{n+1/2}{v_0} z} 
\int_\erre d\xi e^{-iz\xi} e^{\f i 2 \tau \xi^2}
\lan \phi_n , e^{-i \xi \cdot } \phi_0 \ran \widehat V (\xi) e^{-i \xi x } \eta (x -\tau \xi)
\ee
The integral over $z$ can be explicitly computed 
\be
\int_{\erre} dz \, e^{i (\f{n+1/2}{v_0}-\xi) z} = 2\pi \de\lf(\f{n+1/2}{v_0}-\xi\ri) 
\ee
and then one obtains 
\be
\lan \phi_n , \II^{1,+}_1 ( \ve x+ R_0, \cdot ) \ran = 
\f{\sqrt{2\pi}}{i v_0  }\mis e^{\f i 2 \tau \f{(n+1/2)^2}{v_0^2}}
\lan \phi_n , e^{-i \f{n+1/2}{v_0} \cdot } \phi_0 \ran \widehat V \lf(\f{n+1/2}{v_0}\ri) e^{-i \f{n+1/2}{v_0} x } 
\eta \lf(x -\tau \f{n+1/2}{v_0}\ri)
\ee
We also compute the the two terms appearing at the second order. This calculation shows a glimpse
of the difficulties appearing in the general term of the asymptotic expansion which are not hinted 
by the first order term. 
The computation is a bit more delicate and it involves $\delta$, $\delta'$ and principal value distributions.  The first one is the following
\begin{multline}
\lan \phi_n , \II^{2,+}_1 ( \ve x+ R_0, \cdot ) \ran = 
\f{1}{i v_0 \sqrt{2\pi} } \mis\int_{\erre} dz \, e^{i \f{n}{v_0} z} \int_\erre d\xi e^{-iz\xi} e^{\f i 2 \tau \xi^2}
\lan \phi_n , e^{-i \xi \cdot } \phi_0 \ran \widehat V (\xi) e^{-i \xi x }  \\
\cdot \lf[
\eta( x - \tau \xi ) \f i 2 z \xi^2 -z \xi \eta '( x - \tau \xi )
\ri]
\end{multline}
Also in this case the integral over $z$ can be computed explicitly.
\be
\int_{\erre} dz \,z \, e^{i (\f{n+1/2}{v_0}-\xi) z} = \f{2\pi}{i} \de'\lf(\f{n+1/2}{v_0}-\xi\ri) 
\ee
Therefore
\begin{align}
\lan \phi_n , \II^{2,+}_1 ( \ve x+ R_0, \cdot ) \ran 
&= 
 -\f{\sqrt{2\pi}}{ v_0  } \mis\lf\{
i\tau \f{n+\f{1}{2}}{v_0} e^{\f i 2 \tau  \f{(n+\f{1}{2})^2}{v_0^2} }
\lan \phi_n , e^{-i  \f{n+\f{1}{2}}{v_0}  \cdot } \phi_0 \ran \widehat V ( \f{n+\f{1}{2}}{v_0} ) e^{-i  \f{n+\f{1}{2}}{v_0}  x }  \nonumber \ri.\\
&\qquad \cdot \lf[
\eta\lf( x - \tau  \f{n+\f{1}{2}}{v_0}  \ri) \f i 2   \f{(n+\f{1}{2})^2}{v_0^2}  -  \f{n+\f{1}{2}}{v_0}  \eta '\lf( x - \tau  \f{n+\f{1}{2}}{v_0}  \ri)
\ri]\nonumber  \\
&  + \,e^{\f i 2 \tau \f{(n+\f{1}{2})^2}{v_0^2}}
\lan \phi_n , e^{-i \f{n+\f{1}{2}}{v_0} \cdot } \cdot \phi_0 \ran \widehat V ( \f{n+\f{1}{2}}{v_0} ) e^{-i  \f{n+\f{1}{2}}{v_0}  x }  \nonumber \\
&\qquad\cdot\lf[
\eta\lf( x - \tau  \f{n+\f{1}{2}}{v_0}  \ri) \f i 2   \f{(n+\f{1}{2})^2}{v_0^2}  -  \f{n+\f{1}{2}}{v_0}  \eta '\lf( x - \tau  \f{n+\f{1}{2}}{v_0}  \ri)
\ri]\nonumber \\
& +\, e^{\f i 2 \tau \f{(n+\f{1}{2})^2}{v_0^2}}
\lan \phi_n , e^{-i \f{n+\f{1}{2}}{v_0} \cdot }  \phi_0 \ran \widehat V '( \f{n+\f{1}{2}}{v_0} ) e^{-i  \f{n+\f{1}{2}}{v_0}  x }  \nonumber \\
&\qquad\cdot\lf[
\eta\lf( x - \tau  \f{n+\f{1}{2}}{v_0}  \ri) \f i 2   \f{(n+\f{1}{2})^2}{v_0^2}  -  \f{n+\f{1}{2}}{v_0}  \eta '\lf( x - \tau  \f{n+\f{1}{2}}{v_0}  \ri)
\ri]\nonumber \\
& + \,  e^{\f i 2 \tau \f{(n+\f{1}{2})^2}{v_0^2}}
\lan \phi_n , e^{-i \f{n+\f{1}{2}}{v_0} \cdot }  \phi_0 \ran \widehat V ( \f{n+\f{1}{2}}{v_0} )(-ix) e^{-i  \f{n+\f{1}{2}}{v_0}  x }  \nonumber \\
&\qquad \cdot\lf[
\eta\lf( x - \tau  \f{n+\f{1}{2}}{v_0}  \ri) \f i 2   \f{(n+\f{1}{2})^2}{v_0^2}  -  \f{n+\f{1}{2}}{v_0}  \eta '\lf( x - \tau  \f{n+\f{1}{2}}{v_0}  \ri)
\ri]\nonumber \\
& +\,  e^{\f i 2 \tau \f{(n+\f{1}{2})^2}{v_0^2}}
\lan \phi_n , e^{-i \f{n+\f{1}{2}}{v_0} \cdot }  \phi_0 \ran \widehat V ( \f{n+\f{1}{2}}{v_0} )(-ix) e^{-i  \f{n+\f{1}{2}}{v_0}  x } \lf[
\eta\lf( x - \tau  \f{n+\f{1}{2}}{v_0}  \ri)  i \f{n+\f{1}{2}}{v_0}    \nonumber\ri.\\
&\qquad-\lf.\lf(\tau \f i 2   \f{(n+\f{1}{2})^2}{v_0^2}+1 \ri) \eta '\lf( x - \tau  \f{n+\f{1}{2}}{v_0}  \ri)   \lf. 
+ \, \tau  \f{n+\f{1}{2}}{v_0}  \eta ''\lf( x - \tau  \f{n+\f{1}{2}}{v_0}  \ri) \ri]\ri\}
\end{align} 
The second term contains a less trivial integral over $z$.
\begin{multline}
\lan \phi_n , \II^{2,+}_2 ( \ve x+ R_0, \cdot ) \ran = 
\f{1}{i 2 v_0^2 2\pi } \mis \int_{z_1 \leqslant z_2} \!\!\!dz_1 dz_2 \, e^{i \f{n+1/2}{v_0} z_2} \!\! \int_\erre d\xi_1 d\xi_2\, e^{-i(z_1\xi_1+z_2\xi_2)} 
e^{\f i 2 \tau (\xi_1+\xi_2)^2}  \\
\cdot \, \lan \phi_n ,e^{-i \xi_2 \cdot } U\lf( \f{z_1}{v_0} \ri)  e^{-i \xi_1 \cdot } \phi_0 \ran
\widehat V (\xi_1) \widehat V (\xi_2)e^{-i (\xi_1+\xi_2) x } \eta (x -\tau (\xi_1+\xi_2))
\end{multline}
If we expand the harmonic oscillator propagator
\be
U\lf( \f{z_1}{v_0} ; y, z \ri) = \sum_{m=0}^{\infty} e^{-i\f{z_1}{v_0} (m+1/2) } \phi_m (y)  \phi_m (z)
\ee
we are reduced to compute
\be
\int_{z_1 \leqslant z_2} \!\!\! dz_1 dz_2 \, e^{-i\lf(\xi_2- \f{n+1/2}{v_0} \ri)z_2} 
e^{-i\lf( \xi_1+ \f{m+1/2}{v_0}\ri)z_1  } =2\!
\int_{-\infty}^{+\infty} \!\!\! du_1 e^{-i u_1 \lf(\xi_1 +\xi_2 + \f{m-n}{v_0} \ri)}\!\!
\int_0^{\infty}\!\!\! du_2 e^{-i u_2 \lf(\xi_2 -\xi_1 - \f{m+1+n}{v_0} \ri)}
\ee
We have
\begin{align}
\int_{-\infty}^{+\infty} \!\!\! du_1 e^{-i u_1 \lf(\xi_1 +\xi_2 + \f{m-n}{v_0} \ri)}\!\! &=
2\pi \de \lf(\xi_1 +\xi_2 + \f{m+-n}{v_0} \ri) \\
\int_0^{\infty}\!\!\! du_2 e^{-i u_2 \lf(\xi_2 -\xi_1 - \f{m+1+n}{v_0} \ri)}& =
\pi \de \lf(\xi_2 -\xi_1 - \f{m+1+n}{v_0} \ri) - \f{1}{2i\pi} \text{ PV } \f{1}{\xi_2 -\xi_1 - \f{m+1+n}{v_0}}
\label{PV}
\end{align}
We already remarked that the stationary phase expansion is different with respect to the usual one due to 
since 
the stationary point lies on a one dimensional subset of the integration domain boundary. The appearance
of distributions, like the Principal Value in \eqref{PV}, is a manifestation of this fact.

\n 
We have
\begin{multline*}
\lan \phi_n , \II^{2,+}_2 ( \ve x+ R_0, \cdot ) \ran = \f{1}{i 2 v_0^2  }\sum_{m=0}^{\infty}
 \mis e^{-i \f{n-m}{v_0} x }e^{\f i 2 \tau \lf( \f{n-m}{v_0}\ri)^2} \eta \lf(x -\tau \f{n-m}{v_0}\ri)
\int_\erre d\xi_1 d\xi_2\, 
 \de \lf(\xi_1 +\xi_2 + \f{m-n}{v_0} \ri) \\
\lf[
\pi \de \lf(\xi_2 -\xi_1 - \f{m+1+n}{v_0} \ri) - \f{1}{2i\pi} \text{ PV } \f{1}{\xi_2 -\xi_1 - \f{m+1+n}{v_0}}
\ri]  
 \lan \phi_n ,e^{-i \xi_2 \cdot }\phi_m \ran\lan \phi_m  e^{-i \xi_1 \cdot } \phi_0 \ran
\widehat V (\xi_1) \widehat V (\xi_2)
\end{multline*}

\n
We notice that, following the same kind of arguments, in principle one can explicitly  compute also  any other higher order  term of the asymptotic expansion of $\Psi_{\ve}^+(t)$. 

\vs

\n
We want to conclude with a brief discussion of a possible physical application of our result. Let us consider the following initial state 
\ba\label{inst}
&&\Psi_{\ve,0}(R,r) = \alpha_{\ve} \left( \f{e^{i \f{v_0}{\ve^2}R}}{\sqrt{\ve} } \eta(\ve^{-1}(R-R_0))  + \f{e^{-i \f{v_0}{\ve^2}R}}{\sqrt{\ve} } \eta(\ve^{-1}(R+R_0)) \right) \phi_0^{\ve} (r)
\ea
where $v_0 , R_0 >0$, $R_0 <a$ and $\alpha_{\ve}$ is the normalization factor
\ba\label{alp}
&&\alpha_{\ve} = \f{1}{\sqrt{2}} \left( 1 + \int \!\! dx\, \eta(x) \eta(x - 2 R_0 \ve^{-1}) \cos 2 i v_0 \ve^{-1} (x - R_0 \ve^{-1}) \right)^{\!- 1/2}
\ea
We notice that by repeated integration by parts one easily obtains $\alpha_{\ve} = \f{1}{\sqrt{2}} + \mathcal{O}(\ve^k)$, for any $k \in \enne$. In \eqref{inst}  we are assuming that at time zero the test particle is described by a coherent superposition of two wave packets with opposite average momenta and the oscillator is in its ground state. Exploiting theorems 2.2, 2.3, we explicitly  write the first order approximation of the time evolution of \eqref{inst} for a fixed $t>\tau$
\ba\label{timr}
&&\Psi_{\ve}(t) = e^{-i \f{t}{\ve^2}H_0^{\ve}} \Theta_{\ve} + \mathcal{E}_{\ve}(t)
\ea
where $\|\mathcal{E}_{\ve}(t)\| < c \, \ve^2$ and 
\begin{align}
\label{thetar}
&\Theta_{\ve} (R,r) = \alpha_{\ve} \, e^{-2i \f{v_0}{\ve^2}R_0} \sum_{n=0}^{\infty} \Theta_{n,\ve}(R) \, \phi_n^{\ve} (r) \\
&\Theta_{n,\ve}(R)= \f{e^{i \f{v_0}{\ve^2}R}}{\sqrt{\ve} } \eta(\ve^{-1}(R-R_0))  \,\delta_{n0}  + \f{e^{-i \f{v_0}{\ve^2}R}}{\sqrt{\ve} } \eta(\ve^{-1}(R+R_0))  \,\delta_{n0} \nonumber\\
&\qquad\qquad\!+\, \beta_{n,\ve}  \, \sqrt{\ve} \, e^{i\left( \f{v_0}{\ve^2}- \f{n+1/2}{v_0 \ve} \right)R} \eta\big( \ve^{-1}(R\!-\!R_0) \!-\! \tau (n+1/2) v_0^{-1}\big)\\
&\beta_{n,\ve} = \f{\sqrt{2 \pi}}{iv_0} e^{\f{i}{\ve} \left( \f{(n+1/2)R_0}{v_0} + n \tau \right)}  e^{i\f{(n+1/2)^2}{2v_0^2}} \lan\phi_n, e^{-i\f{n+1/2}{v_0} \cdot} \phi_0 \ran \widehat{V} \big( v_0^{-1} (n+1/2) \big) \label{beta}
\end{align}
\n
It is also convenient to consider  the Fourier transform $\widehat{\Theta}_{n,\ve}$ of $\Theta_{n,\ve}$, which is explicitly given by
\begin{multline}
\label{thetak}
\widehat{\Theta}_{n,\ve}(K) = \! \sqrt{\ve} \, e^{-i \left(K-\f{v_0}{\ve^2} \right) R_0} \,\hat{\eta}(\ve K \!-\! \ve^{-1} v_0) \, \delta_{n0} + \sqrt{\ve} \, e^{i \left(K+\f{v_0}{\ve^2} \right) R_0} \,\hat{\eta}(\ve K \!+\! \ve^{-1} v_0) \, \delta_{n0} \\
+\, \beta_{n,\ve} \, \ve^{3/2} \, e^{-i \left( K - \f{v_0}{\ve^2} + \f{n+1/2}{v_0 \ve}\right) \left( \f{\ve \tau (n+1/2)}{v_0} +R_0 \right)} \hat{\eta} \big(\ve K \!-\! \ve^{-1} v_0 \!+\! v_0^{-1}(n\!+\!1/2) \big) 
\end{multline}
Starting from \eqref{timr},  our aim is to compute approximate expressions for the probabilities of the outcomes of the measurement of some interesting observables relative to the system which could possibly be performed for $t>\tau$ . In particular we are interested in:

\n
- the probability to find the energy $E_0$ or $E_1$ for the oscillator, denoted by $\mathcal{P}_{0}$,  $\mathcal{P}_{1}$ respectively;

\n

\n
- the probability to find a positive momentum for the test particle and the energy $E_0$ or $E_1$ for the oscillator, denoted by 
$\mathcal{P}_{+,0}$ and $\mathcal{P}_{+,1}$ respectively. 

\n
In order to control the approximations, we observe that for any projection operator $P$ in $L^2(\erre^2)$, which commutes with $H_0^{\ve}$, one has
\ba\label{appro}
&&\left| (\Psi_{\ve}(t), P \Psi_{\ve}(t)) - (\Theta_{\ve},P \Theta_{\ve}) \right| \leq \|\mathcal{E}_{\ve}(t) \| \Big( 2 \|P \Theta_{\ve} \| + \|\mathcal{E}_{\ve}(t)\|\Big) \leq c \, \ve^2 \Big( \|P \Theta_{\ve} \| + \ve^2 \Big)
\ea

\n
From \eqref{timr}, a direct application of the Born's rule and \eqref{appro}  we obtain
\ba\label{p0}
&&\mathcal{P}_0= |\alpha_{\ve}|^2 \int\!\!dR\, |\Theta_{0,\ve}(R)|^2 + \mathcal{O}(\ve^2) = 1 +\mathcal{O}(\ve)\\
&&\mathcal{P}_1 = |\alpha_{\ve}|^2 \int\!\!dR\, |\Theta_{1,\ve}(R)|^2 + \mathcal{O}(\ve^3) = \f{1}{2}\,  |\beta_{1,\ve}|^2\, \ve^2 + \mathcal{O}(\ve^3)\label{p1}
\ea
Moreover
\ba\label{p+0}
&&\mathcal{P}_{+,0}= |\alpha_{\ve}|^2 \int_0^{\infty} \!\!\!dK\, |\widehat{\Theta}_{0,\ve} (K)|^2  +\mathcal{O}(\ve^2) = \f{1}{2}\, + \mathcal{O}(\ve)\\
&&\mathcal{P}_{+,1}= |\alpha_{\ve}|^2 \int_0^{\infty} \!\!\!dK\, |\widehat{\Theta}_{1,\ve} (K)|^2  +\mathcal{O}(\ve^3)=\f{1}{2}\,  |\beta_{1,\ve}|^2\, \ve^2 + \mathcal{O}(\ve^3)
\ea
Hence we have
\ba\label{p+00}
&&\f{\mathcal{P}_{+,0}}{\mathcal{P}_0}= \f{1}{2} \, +\mathcal{O}(\ve)\\
&&\f{\mathcal{P}_{+,1}}{\mathcal{P}_1}= 1 \, +\mathcal{O}(\ve)\label{p+11}
\ea
The l.h.s. of formulas \eqref{p+00}, \eqref{p+11} could be interpreted as a sort of conditional probabilities. More precisely, \eqref{p+00} says that if the oscillator remains in the ground state then the probability to find a positive momentum for the test particle is approximately one-half. This essentially means that in this case the superposition state for the test particle survives.  On the other hand, \eqref{p+11} says that if the oscillator is in the first excited state then the probability to find a positive momentum for the test particle is approximately one. In such case the test particle is described by a wave packet with positive momentum moving to the right of the oscillator.  
It should be stressed that both "histories" are contained in the complete wave function of the system and a choice is made only when a real  measurement of the energy of the oscillator is performed.


\section{Appendix}
\n
In this appendix we give the proof of proposition \ref{prop1} which is essentially based on 
algebraic manipulations and the use of  Fourier transform. From (\ref{il}) we have

\begin{multline}
\lan \phi_n , I_l (t) \ran =(-i)^l  \sum_{n_1 \ldots n_l=0}^{\infty} \delta_{n_1 0} 
\int_0^t ds_l  \int_0^{s_l}  ds_{l-1} \ldots  \int_0^{s_2}  ds_1 
\lan \phi_n , e^{i\f{s_l}{\ve^2} H_0} V e^{-i\f{s_l}{\ve^2} H_0} \phi_{n_l} \ran\\
\cdot \lan \phi_{n_l} , e^{i\f{s_{l-1}}{\ve^2} H_0} V e^{-i\f{s_{l-1}}{\ve^2} H_0} \phi_{n_{l-1}} \ran
\ldots 
\lan \phi_{n_2} , e^{i\f{s_{1}}{\ve^2} H_0} V e^{-i\f{s_{1}}{\ve^2} H_0} \phi_{n_{1}} \ran
\psi^{+}
\label{buta}
\end{multline}
For the generic scalar product appearing in the integrand in  \eqref{buta} we have
\be\label{butatu}
\lan \phi_{n} , e^{i\f{s}{\ve^2} H_0} V e^{-i\f{s}{\ve^2} H_0} \phi_{m} \ran  =
e^{\f{i}{\ve}(n-m)s}
e^{i\f{s}{\ve^2} h_0} V_{nm} e^{-i\f{s}{\ve^2} h_0}   
\ee
where
\begin{align}\label{butati}
V_{nm}^{\ve}(R)  &=\f{1}{\ve} \int dr\, \phi_n \lf( \ve^{-1}(r-a) \ri)
V \lf( \ve^{-1}(R-r) \ri) \phi_m \lf( \ve^{-1}(r-a)\ri) \nonumber \\
&  = \int \!\! dx \; \phi_n(x) \phi_m(x) V\left( \ve^{-1}(R-a) - x \right) \nonumber\\
& = \int \!\! d \xi \; \widehat{V}(\xi) \widehat{\phi_n \phi_m}(\xi) \, e^{-i \xi \f{R-a}{\ve}} 
\end{align}
Furthermore for any $g \in L^2(\erre)$ we have
\ba\label{pseu}
&&\left( e^{i \f{s}{\ve^2} h_0} e^{-i \xi \frac{\cdot}{\ve} } e^{-i \f{s}{\ve^2} h_0} g\right)(R)  =
\frac{1}{\sqrt{2\pi} }\int \!\! dk\, e^{i kR}  \lf(
e^{i \f{s}{\ve^2} h_0} e^{-i \xi \frac{\cdot}{\ve} } e^{-i \f{s}{\ve^2} h_0} g
\ri)^{\!\widehat{}} \!(k) \nonumber\\
& & \qquad =\frac{1}{\sqrt{2\pi} }\int \!\! dk\, e^{i kR} e^{i\frac{k^2 \ve^2}{2} s}   \lf(
  e^{-i \xi \frac{\cdot}{\ve} } e^{-i \f{s}{\ve^2}h_0} g\ri)^{\!\widehat{}} \! (k) \nonumber\\
& &  \qquad=\frac{1}{\sqrt{2\pi} }\int \!\!dk\, e^{i kR} e^{i\frac{k^2 \ve^2}{2} s}   \lf(
   e^{-i \f{s}{\ve^2} h_0} g \ri)^{\!\widehat{}}\! \lf(k+ \ve^{-1} \xi \ri) \nonumber\\
& & \qquad =\frac{1}{\sqrt{2\pi} }\int \!\!dk\, e^{i kR} e^{i\frac{k^2 \ve^2}{2} s}  
e^{-i\frac{s \ve^2}{2} \lf(k+ \f{\xi}{\ve} \ri)^2 } \widehat{g}\lf(  k+ \ve^{-1}\xi \ri) \nonumber\\
& &  \qquad=\frac{1}{\sqrt{2\pi} }e^{-is\frac{\xi^2}{2} }\int \!\!dk\, e^{i kR}   e^{-i\ve s k \xi} 
  \widehat{g}\lf( k+ \ve^{-1} \xi \ri) \nonumber\\
  &&\qquad=e^{i s \f{\xi^2}{2}} \, e^{-i \f{R}{\ve} \xi} \, g ( R - \ve s \xi)
\ea
From \eqref{butatu}, \eqref{butati}, \eqref{pseu} we conclude 
\be
\left(\!\lan \phi_{n} , e^{i\f{s}{\ve^2} H_0} V e^{-i\f{s}{\ve^2} H_0} \phi_{m} \ran g \!\right)\!(R) =e^{\f{i}{\ve}(n-m)s}\!
\int \!\! d \xi \; \widehat{V}(\xi) \widehat{\phi_n \phi_m}(\xi) \, e^{i \xi \f{a}{\ve}} 
e^{i s \f{\xi^2}{2}} \, e^{-i \f{R}{\ve} \xi} \, g ( R - \ve s \xi)
\label{yoru}
\ee
Substituting \eqref{yoru} into \eqref{buta} and taking into account the explicit expression of $\psi^+$
we obtain \eqref{pnil}.

\vs
\hfill $\Box$

\vs
\vs

\vs
\vs

\end{document}